\documentclass[
    letterpaper,
    10pt,
    aps,
    pra,
    twocolumn,
    floatfix,
    superscriptaddress,
    showkeys
]{revtex4-1}

\usepackage[english]{babel}
\usepackage[T1]{fontenc}
\usepackage[utf8]{inputenc}

\usepackage{amsmath}                 
\usepackage{amssymb}                 
\usepackage{physics}                 
\usepackage{array}                   
\usepackage{xfrac}                   
\usepackage{graphicx}                
\usepackage{xcolor}                  
\usepackage[
    subrefformat=parens,
    labelformat=parens,
    caption=false
]{subfig}                            
\usepackage[export]{adjustbox}       
\usepackage{csvsimple}               
\usepackage{siunitx}                 
\usepackage{booktabs}                
\usepackage{xr}                      
\usepackage{multirow}                
\usepackage{enumitem}                
\usepackage{rotating}                
\usepackage{tikz}                    
\usepackage{makecell}                
\usepackage{hyperref}                

\colorlet{refcolor}{blue!80!black}
\hypersetup{
    colorlinks,
    linkcolor=refcolor,
    citecolor=refcolor,
    urlcolor=refcolor
}

\newcommand{\longcomment}[1]{}

\newcolumntype{x}[1]{>{\centering\let\newline\\\arraybackslash\hspace{0pt}}p{#1}}

\newcommand{\taaast}{\textasteriskcentered\textasteriskcentered\textasteriskcentered}
\newcommand{\taast}{\textasteriskcentered\textasteriskcentered}
\newcommand{\tast}{\textasteriskcentered}

\begin{document}

\title{Investigating institutional influence on graduate program admissions by modelling physics GRE cut-off scores}

\author{Nils J. \surname{Mikkelsen}}
\affiliation{Center for Computing in Science Education \& Department of Physics, University of Oslo, N-0316 Oslo, Norway}
\author{Nicholas T. \surname{Young}}
\affiliation{Department of Physics and Astronomy, Michigan State University, East Lansing, Michigan 48824}
\affiliation{Department of Computational Mathematics, Science, and Engineering, Michigan State University, East Lansing, Michigan 48824}
\author{Marcos D. \surname{Caballero}}
\email[Corresponding Author: ]{caballero@pa.msu.edu}
\affiliation{Center for Computing in Science Education \& Department of Physics, University of Oslo, N-0316 Oslo, Norway}
\affiliation{Department of Physics and Astronomy, Michigan State University, East Lansing, Michigan 48824}
\affiliation{Department of Computational Mathematics, Science, and Engineering, Michigan State University, East Lansing, Michigan 48824}
\affiliation{CREATE for STEM Institute, Michigan State University, East Lansing, Michigan 48824}

\date{\today}

\begin{abstract}
    Despite limiting access to applicants from underrepresented racial and ethnic groups, the practice of using hard or soft GRE cut-off scores in physics graduate program admissions is still a popular method for reducing the pool of applicants. The present study considers whether the undergraduate institutions of applicants have any influence on the admissions process by modelling a physics GRE cut-off score with application data from admissions offices of five universities. Two distinct approaches based on inferential and predictive modelling are conducted. While there is some disagreement regarding the relative importance between features, the two approaches largely agree that including institutional information significantly aids the analysis. Both models identify cases where the institutional effects are comparable to factors of known importance such as gender and undergraduate GPA. As the results are stable across many cut-off scores, we advocate against the practice of employing physics GRE cut-off scores in admissions.
\end{abstract}
\keywords{Physics graduate admissions, physics GRE, institutional influence, logistic regression modelling, supervised machine learning.}

\maketitle

\section{Introduction}
While recent studies have called into question the over-reliance on Graduate Record Examination (GRE) scores in physics graduate admissions \cite{miller_test_2014,miller_typical_2019}, filtering applicants based on a strict or effective minimum score is still a popular practice today \cite{potvin_investigating_2017}. Given the role of the GRE in admissions, understanding the factors influencing GRE scores may provide insight into how, when compared to other science, technology, engineering and mathematics (STEM) disciplines, the physics graduate admissions process has failed to improve gender, racial, and ethnic diversity by systematically excluding these applicants \cite{porter_women_2019,laura_merner_african_2017}.

A number of studies have investigated correlations between GRE scores and demographics \cite{miller_test_2014,miller_typical_2019}, but little attention has been given to the institutional backgrounds of applicants. An applicant's undergraduate background could play a significant role in their graduate application \cite{young_physics_2020}. Institutions offering a PhD program themselves would likely place more emphasis on both preparing and motivating undergraduate students for further studies. Larger physics departments with more resources are able to offer students more advanced course-work and hands-on experimental work as well as provide a larger variety of staff expertise. Larger undergraduate programs can facilitate network-building, both between students and faculty members, and collaboration via projects and study-groups. Although attributes such as motivation and opportunity cannot be appropriately measured, their effects on the GRE can be linked to metrics such as the size and type of institutions as was done in Halley et al. \cite{halley_graduate_1991}. In order to estimate these institutional effects, we have analyzed the Physics GRE Subject Test (P-GRE) scores of graduate program applications from four public universities and one private university.

The applications include a variety of information, but the present study will focus on numerical and categorical data, all of which constitutes a mixture of data structures. A number of recent studies working with similar data have approached the problem using machine learning methods \cite{young_using_2019,zabriskie_using_2019}.
\longcomment{Many machine learning algorithms provide an innate flexibility that lends itself to problems with mixed data, albeit losing the interpretative nature of more conventional modelling methods in the process.}
Many machine learning methods lend themselves to problems with mixed data, albeit they do not share the interpretability of more conventional modelling methods. The present study will employ both approaches, contrasting and comparing the results.

The aim of this study is to continue the discussion on the practice of employing formal or informal P-GRE score cut-offs in graduate admissions using a combination of modelling and machine learning methods. The idea is to analyze the P-GRE scores of PhD program applicants with respect to applicants' undergraduate Grade Point Average (U-GPA), demographics and institutional background. Our guiding research questions (RQs) are as follows.
\begin{enumerate}
    \item To what extent does the applicant's undergraduate institution influence whether they are able to attain a minimum P-GRE score expected by an admissions committee?
    \item To what degree do the institutional effects compare to known effects such as U-GPA, gender and race?
    \item How do the results depend on the specific cut-off chosen by the admissions office?
    \item How well do the conventional and machine learning approaches agree on RQs 1, 2 and 3?
\end{enumerate}
\section{Background}
Following the calls for increasing diversity in STEM disciplines, there has been a steady growth of women's and ethnic/racial minorities' representation over the past couple of decades \cite{ivie_beyond_2018}. Despite the progress however, physics has seen particularly poor development in comparison. Since the late 1990s, the percentage of bachelor and PhD degrees awarded to women in physics has stagnated at about 20\%, mirroring similar numbers of engineering and computer science \cite{porter_women_2019}. The numbers are even more concerning for racial minorities who during the three-year period 2014-2016 earned 11\% of bachelor degrees and only 7\% of PhD degrees \cite{ivie_beyond_2018}. The discrepancy in female, racial and ethnic representation likely stems from variety of factors involving admission and retention issues, many of which are rooted in cultural and structural problems including sexual harassment and systemic racism \cite{aycock_sexual_2019, rosa_educational_2016,hyater-adams_critical_2018}.

In her extensive review of the general practices of graduate program admissions, \emph{Inside Graduate Admissions} (2016) \cite{posselt_inside_2016}, Posselt notes that most admissions (in the natural sciences as well as in the humanities and social sciences) measured students' merit primarily on the basis of their undergraduate GPA (U-GPA) and GRE scores alone. Indeed, Young and Caballero were able to predict the admittance of prospective physics PhD students with 75\% accuracy using machine learning methods based only on their U-GPA and P-GRE score \cite{young_using_2019}. The GRE test makers, Educational Testing Service (ETS), recommends against the use of GRE scores as the sole basis for admissions decisions, particularly emphasizing the practice of filtering applicants based on a minimum cut-off score \cite{educational_testing_service_guide_2019}. Despite this, Potvin et al. found that 32\% of physics graduate program admissions state they filter applicants with a minimum P-GRE score \cite{potvin_investigating_2017}. Furthermore, of the programs that say they do not filter applicants, several reported using a ``\emph{rough cut-off}'' or wanting a ``\emph{preferable score}'', suggesting that more than 32\% of programs filter applicants in practice.

As highlighted by Miller and Stassun in 2014 \cite{miller_test_2014}, on average, women score 80 pts lower than men on the GRE in the physical sciences, while Black test-takers score 200 pts lower than white test-takers. The authors further note that the practice of filtering prospective students with a minimum score, which is in violation with ETS's own guidelines, thus ``\emph{adversely effects women and minority applicants}''. In addition to limiting access for minority applicants during the application process, the GRE also acts as barrier to apply. In a survey of prospective students from underrepresented racial and ethnic groups interested in pursuing a PhD in physics who ultimately chose not to apply, Cochran et al. notes that the GRE was the ``\emph{most common theme}'' expressed by students as a barrier to apply \cite{cochran_identifying_2018}.

In spite of its established popularity in admissions, the GRE's ability to identify promising students has recently been called into question. One study found that while requiring a minimum P-GRE score limits access to physics graduate program applicants from minority groups, GRE scores were incapable of predicting PhD completion \cite{miller_typical_2019}. In a 2015 survey of prize-winning postdoctoral fellows in astronomy \cite{levesque_physics_2015}, Levesque et al. found that the P-GRE scores of fellows did not adhere to any minimum percentile score, suggesting that the GRE is also a poor estimator for future research excellence. The authors further point out that a minimum percentile score of 60\% would have eliminated 44\% of participants, including 60\% of female fellows. The inability of the GRE to identify promising students has also been noticed by other groups such as the National Science Foundation, which recently decided to drop the GRE from the application to their Graduate Research Fellowship Program (see FAQ no. 52 \cite{national_science_foundation_frequently_2020}).
\longcomment{Prior work has primarily focused on how using the GRE in admissions limits access to underrepresented groups, in addition to its poor efficacy for predicting graduate program completion. In a 1991 study, Halley et al. investigated how the topics covered by P-GRE compared with physics major curriculum by analyzing the P-GRE scores of students from different institution types \cite{halley_graduate_1991}. The authors note that students from "top" institutions had a higher portion of correct answers, and among the students from the top institutions, those from institutions with graduate programs had the highest portion of correct answers. However, this study is both nearly 30 years old and worked with an imbalanced sample (701 test-takers in total, 21 of which attended top undergraduate institutions). Since then, the GRE has evolved and the number of physics degrees awarded annually has almost doubled \cite{mulvey_physics_2015}. Nowadays, the GRE does not penalize incorrect answers, i.e., guessing, which has likely changed the way students approach the test. To our knowledge, there has not been a modern study incorporating institutional effects of how the P-GRE relates to graduate admissions in physics.}

Prior work has typically focused on admissions committees' over-reliance on the GRE and the consequences of using cut-off scores in graduate admissions \cite{attiyeh_testing_1997,miller_test_2014,miller_typical_2019,posselt_metrics_2019}. Missing from the conversation is an understanding of what institutional factors, which come into play during applicants' undergraduate study (or even earlier), may influence GRE scores. In a 1991 study, Halley et al. investigated how the topics covered by P-GRE compared with physics major curriculum by analyzing the P-GRE scores of students from different institution types \cite{halley_graduate_1991}. The authors noted that the portion of correct answers was higher for students from "top" institutions, and highest for students from "top" institutions with graduate programs. However, this study is both nearly 30 years old and worked with an imbalanced sample (701 test-takers in total, 21 of which attended a top undergraduate institution). Since then, the GRE has evolved and the number of physics degrees awarded annually has almost doubled \cite{mulvey_physics_2015}. Nowadays, the GRE does not penalize incorrect answers, i.e., guessing, which has likely changed the way students approach the test. To our knowledge, there has not yet been a modern study analyzing how institutional factors may affect GRE scores.

\section{Methods}
The target for this investigation is to explain whether a student scores \emph{above} or \emph{below} a P-GRE cut-off score selected by an admissions committee. This is encoded using a binary response variable named \texttt{ABOVE} with the interpretation that an applicant with a score \emph{above or equal to} the cut-off has \(\texttt{ABOVE}=1\), and an applicant with a score \emph{below} the cut-off has \(\texttt{ABOVE}=0\). That is, given a test score of \(x\) and a cut-off score of \(C\), we define
\begin{equation}
\label{eq:ABOVE_definition}
    \texttt{ABOVE}=\begin{cases}
    1,&x\geq C,\\
    0,&x<C.
    \end{cases}    
\end{equation}
The reader should recall that the possible scores on GRE subject tests range from 200 to 990 in 10 pt. intervals. We have focused on P-GRE cut-off scores ranging from 620 to 800 pt., corresponding to the 32nd and 67th national percentiles \cite{educational_testing_service_gre_2019}. Typical P-GRE cut-off scores lie in the region of 700 \cite{miller_typical_2019}.

The data used in this study consists of 2017/2018 admissions records for physics graduate programs from 4 public universities in the Big Ten Academic Alliance and one private Midwestern university. The records contain unidentified profiles of program applicants with information regarding their GRE performance, undergraduate GPA, ethnicity and race, gender, etc. In addition, the records also include which institution the applicants attended during their bachelor's degrees. Complementary data describing the bachelor-institutions have been added from three sources: the 2015 Carnegie Classification of Institutions of Higher Education \cite{center_for_postsecondary_research_carnegie_2016}, Barron's selectivity index \cite{barrons_educational_series_inc_college_division_barrons_nodate}, and 2017-2018 surveys of American universities by the American Institute of Physics (AIP) \cite{nicholson_roster_2018,nicholson_roster_2019}. The additional data describes several aspects of the institutions such as institution-wide admissions selectivity and the size of physics programs. The main idea is to study the statistical effects from applicants' institutional backgrounds using this complementary data.
\subsection{On the data}\label{sec:on_the_data}
\begin{figure}
    \centering
    \includegraphics[width=0.48\textwidth]{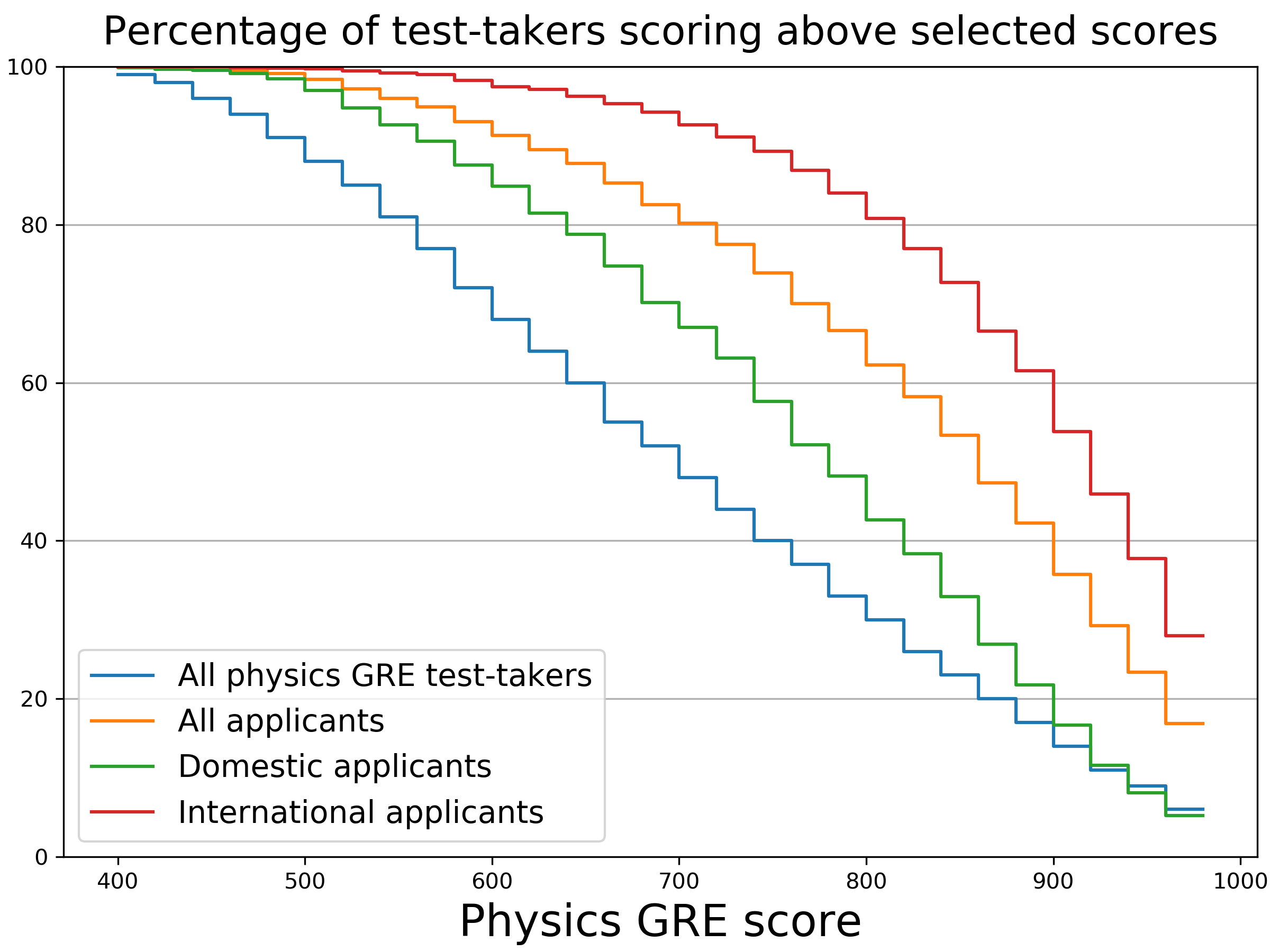}
    \caption{A comparison of the P-GRE distribution between national data \cite{educational_testing_service_gre_2019} and data used in this study. The analysis is primarily concerned with domestic applicants (green curve).}
    \label{fig:PGRE_distribution}
\end{figure}

The admissions records contain 5738 applications in total, but only 5314 (ca. 93\%) of them include the students' P-GRE scores. Applications without P-GRE scores are ignored to avoid influencing the P-GRE distribution. Of the remaining applications, 2575 are domestic (ca. 48\%). This study will focus entirely on domestic students for two main reasons. First, the P-GRE distribution for international students is much more saturated with perfect scores than the distribution is for domestic students. The saturation problem is visualized in figure \ref{fig:PGRE_distribution}: The percentage of international students scoring above the selected cut-off scores both starts off much higher and falls off much slower than the percentage of domestic students. Second, because there is not a systematic collection of graduation records for non-US schools, it is difficult to reliably collect the necessary information from every international student.

Because the applicants are not identified, several applications may come from the same student. While these applications are unique in the sense that each application addresses a different school, they count as duplicated applications in this analysis by virtue of being from the same student. Duplicate applications could have an effect on the results, most notably in the logistic regression model that relies on independent observations (see supplementary material). By comparing applications according to demographics and academic performance, a number of possible duplicate applications have been identified. In case all candidates are duplicates, roughly 18.8\% of the data should be ignored. Because the applications are anonymous, i.e., it is impossible to determine whether two applications belong to the same student, we will conduct the analysis both with and without the possible duplicates.
\subsubsection{The raw features}
In addition to the P-GRE score, thirteen features, or variables, have been selected for analysis. A summary of the features and their sources is given in Table \ref{tab:feature_summary}.

\begin{table}[b]
    \centering
    \caption{A summary of the features used in this study.}
    \label{tab:feature_summary}
    \begin{tabular}{lcc}
        \toprule
        Feature                      & Type        & Source     \\
        \midrule
        Physics GRE score             & continuous  & Admissions \\
        Undergraduate GPA             & continuous  & Admissions \\
        Gender                        & binary      & Admissions \\
        Race                          & categorical & Admissions \\
        Carnegie Classification       & categorical & Carnegie   \\
        Undergrad Population Profile  & categorical & Carnegie   \\
        Funding category              & categorical & Carnegie   \\
        ACT selectivity category      & categorical & Carnegie   \\
        Minority Serving Institution  & binary      & Carnegie   \\
        Barron's selectivity index    & categorical & Barron's   \\
        No. bachelor graduates (2017) & continuous  & AIP Survey \\
        No. bachelor graduates (2018) & continuous  & AIP Survey \\
        No. PhD graduates (2017)      & continuous  & AIP Survey \\
        No. PhD graduates (2018)      & continuous  & AIP Survey \\
        \bottomrule
    \end{tabular}
\end{table}

The features from the admissions records include the applicants' P-GRE score, U-GPA, gender, and race. Note that the gender feature is encoded as a binary variable; while we acknowledge that gender is not binary, more detailed descriptions were not collected by the admissions offices \cite{traxler_enriching_2016}. Similarly, different practices regarding the collection of data on racial and ethnic backgrounds has limited the scope of the race feature. See Posselt et al. for more details regarding collection of data on racial and ethnic backgrounds by admissions offices \cite{posselt_metrics_2019}. The features from the admissions records constitute the applicant-specific component of the models, while the remaining features comprise the institutional component.
 
Of the Carnegie features, the two most prominent are the (2015) Carnegie (basic) classification of institutions and the (2015) undergraduate population profile classification. The basic classification is an overall categorization of the academic degrees offered and awarded by the institutions, e.g. \emph{Doctoral university with high research activity} and \emph{Master's college with large programs}. The undergraduate population profile classification characterizes the typical undergraduate population according to three metrics: portion of full-time undergraduates, academic achievements of first-year and first-time students, portion of entering transfer students. In addition, the Carnegie features also include the institutions' Funding category and ACT selectivity category, and whether the institutions are Minority Serving Institutions (MSI). The ACT category measures the entry selectivity of admissions offices by grouping all institutions according to the ACT scores of first-year bachelor students, and MSI indicates whether an institution satisfies the requirements for a Minority Serving Institution \cite{us_department_of_education_lists_nodate}.

Lastly, Barron's provides the Profile of American Colleges \cite{barrons_educational_series_inc_college_division_barrons_nodate}, which is an index for institution-wide admissions selectivity, and the AIP surveys provide the numbers of bachelor and PhD students graduating in physics.

The data will be analyzed using two different data analysis methods based on logistic regression modelling and predictive machine learning analysis (described in Sec. \ref{sec:methods_data_analysis}). As they stand, the raw features are not well-suited for logistic regression due to computational issues as well as modelling-related difficulties. The remaining part of this section describes our data preprocessing and modelling choices. See Sec. \ref{sec:data_processing_modelling_choices} for a discussion of potential issues. Because the predictive analysis requires less preprocessing than logistic regression, we provide a summary of all the models used in this study in Sec. \ref{sec:clarifying_models} to avoid confusion.
\subsubsection{Underrepresented racial and ethnic minorities}
\begin{figure}[th]
    \centering
    \includegraphics[width=0.47\textwidth]{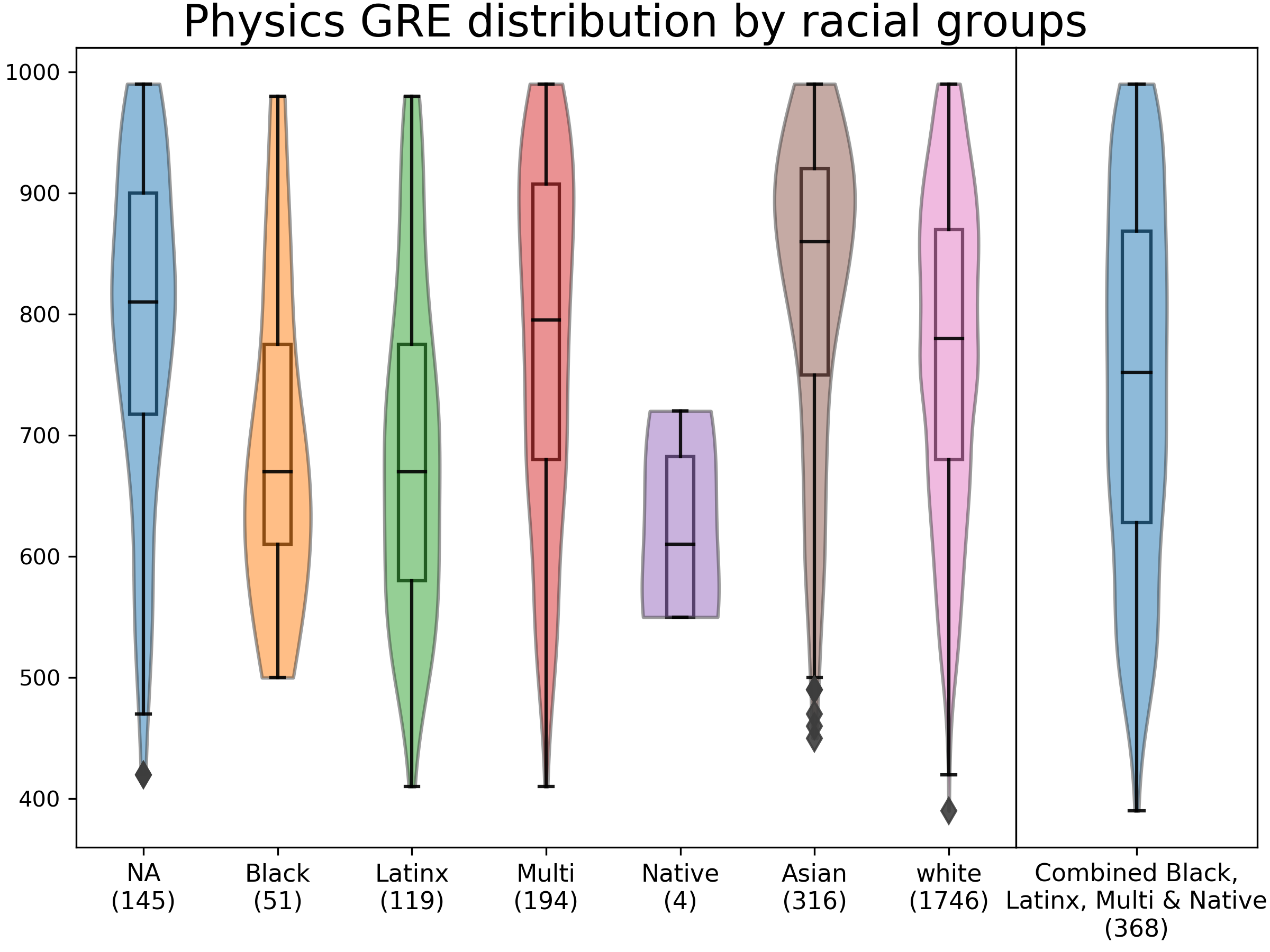}
    \caption{Estimated P-GRE distributions by racial and ethnic groups (number of applicants indicated in parenthesis). Note that the combined distribution normalizes the differences between the combined groups.}
    \label{fig:PGRE_violin_by_race}
\end{figure}

The small representation seen of applicants from racial and ethnic minorities (Black, Latinx, Multi and Native) is of computational concern because logistic regression fairs poorly with low-frequency categories \cite{hosmer_applied_2000}. Because initial tests including every racial group produced results with limited statistical power (e.g. infinite \(p\)-value confidence intervals), we combined racial and ethnic minorities in an underrepresented minority (URM) category despite Teranishi's warning \cite{teranishi_race_2007}. This also combines their P-GRE distributions (see Figure \ref{fig:PGRE_violin_by_race}), leading to loss of information. This issue is further discussed in Sec. \ref{sec:data_processing_modelling_choices}.
\vspace{-0.5mm} 
\subsubsection{The Carnegie classification \& undergraduate population profile}
While the Carnegie classification and undergraduate population profile support 34 and 16 unique categories each, the limited pool of applications leaves many categories empty or with only a handful of applicants. Most of the categories are difficult to combine into meaningful groups. Thus, to avoid computational issues the features are replaced by the binary labels: \emph{Doctoral university w/ highest research activity} and \emph{Most selective undergraduate population}.
\subsubsection{Funding category \& ACT selectivity category}
Similar to the Carnegie features, both \emph{Funding category} and \emph{ACT selectivity category} have categories with too few applicants. To avoid complications, the features are reduced to binary labels \emph{Public Funding} and \emph{Most ACT-selective}, which, respectively, indicate whether the institution is publicly funded and if the institution is in the most selective ACT category.
\subsubsection{Barron's selectivity index}
Barron's selectivity index is an admissions selectivity measure that categorizes institutions according to school competitiveness. In decreasing order of competitiveness, the categories include \emph{most competitive}, \emph{highly competitive}, \emph{very competitive}, \emph{competitive}, \emph{less competitive} and \emph{non-competitive}. Additional ``plus'' categories such as \emph{highly competitive plus} have been collapsed into their corresponding ordinary levels. In this study, admissions selectivity is used as a metric for an institution's resources and staff experience. Because admissions selectivity is expected to have an effect only for the most selective schools, the selectivity categories less competitive than \emph{most} and \emph{highly competitive} are combined to a \emph{not as competitive} category.
\subsubsection{No. bachelor/PhD graduates 2017/2018}\label{sec:no_bach_phd_grads}
In this study, the AIP features (see Table \ref{tab:feature_summary}) provide a measure of the size of undergraduate physics departments. As larger departments typically have more financial resources available and may offer students more opportunities for advanced coursework or research, the P-GRE scores of applicants from larger programs is expected to be higher \cite{halley_graduate_1991}. However, because of the variety of institutions and physics programs, a systemic effect is expected to only emerge for very large physics programs. Instead of analyzing the raw number of graduates, a physics program is classified as \emph{large} if the number of graduates is above the 75th national percentile \cite{mulvey_physics_2015}.

While the typical size of physics departments is unlikely to change on a yearly basis for most institutions, the exact number of graduates is much more sensitive to variation. Moreover, the applicants spent several years at the undergraduate institutions, thus it is unreasonable to estimate the general size of the physics departments using data from a single year. Because the statistical models cannot include data on both years simultaneously (i.e., as individual features) due to correlation issues, the 2017 and 2018 data must be combined (bachelor and PhD features separated). For most institutions, the difference in the number of bachelor/PhD graduates between 2017 and 2018 is not significant enough to have any effect on the analysis. However, because the difference is large for some institutions, naively selecting, say, the average could overestimate or underestimate the size of some departments. In addition, there are some institutions for which data is missing for either 2017 or 2018. To avoid inaccurate single-point estimates of department sizes, the maximum and minimum cases are considered separately. In the \emph{maximum graduates models}, the maximum number of bachelor and PhD students between the 2017 and 2018 data is included, and vice versa in the \emph{minimum graduates models}. For institutions with missing data, any available data is used for both models.
\subsection{Methods for data analysis}\label{sec:methods_data_analysis}
The following section provides a brief overview of the methods used in this study. Additional details are provided as supplementary material. Because logistic regression is likely familiar to a greater audience, more time is spent on the machine learning methods.
\subsubsection{Logistic regression modelling}
Logistic regression analysis is a technique for modelling a binary response \(y\in\{0,1\}\) with respect to explanatory variables \(x_1\ldots,x_k\), which may consist of a mixture of continuous and discrete data. While binary data is naturally handled by logistic regression, categorical (discrete) data with \(M>2\) no. categories must be encoded using \(M-1\) binary variables according the \emph{one-hot encoding} scheme (see supplementary material for details). The response is modelled according to the odds equation,
\begin{equation}
\label{eq:odds_equation}
\text{odds}(p)=\exp\big(\beta_1x_1+\cdots+\beta_k x_k+\epsilon\big),
\end{equation}
where \(p\) is the probability of the outcome \(y=1\), \(\beta_i\) is the regression coefficient of \(x_i\) and \(\epsilon\) is an error term. The regression coefficients are determined numerically using an iterative scheme based on maximum likelihood estimation. In our study this is handled by the \texttt{glm} function in \texttt{R} \cite{friedman_regularization_2010}

A major benefit of logistic regression modelling is the interpretability of its regression coefficients. When \(x_i\) increases by 1 unit, the odds change by a factor of \(\exp(\beta_i)\) called the odds ratio:
\begin{equation}
    \label{eq:odds_ratio}
    \text{OR}(p;x_i)=\frac{\text{odds}(p;x_i+1)}{\text{odds}(p;x_i)}=e^{\beta_i}.
\end{equation}
The interpretation of the odds ratio depends on whether \(x_i\) is continuous or categorical. For continuous features, the change is associated with a unit increase in \(x_i\). For binary features, the change is associated with a switch in \(x_i\) from category 0 to category 1. Because multi-leveled categorical features are encoded with binary features, each binary represents a change from the reference category to the category associated with the binary. Odds ratios below 1 are inverted so that \(1/\text{OR}(x_i)\) is the odds ratio associated with a unit decrease in \(x_i\) or a switch in \(x_i\) from category 1 to category 0. In order to avoid interpretation issues relating to very large or very small continuous features, it is customary to standardize continuous features by centering the mean about 0 and normalizing the variance to 1. For standardized features, the odds ratio is associated with an increase in the original feature by one standard deviation.

Alongside the regression coefficients, the \texttt{glm} function provides the corresponding \(p\)-values. To avoid multiple comparisons problems, the \(p\)-values are adjusted according to the Bonferroni correction. For a logistic regression model with \(N\) features, the Bonferroni-adjusted \(p\)-value is \(\tilde{p}=pN\). We follow common practice and include three levels of significance: \(\alpha=0.05\), \(\alpha=0.01\) and \(\alpha=0.001\).

Because logistic regression is unable to handle missing values, we follow Nissen et al.'s recommendation of imputing the missing data instead of discarding it \cite{nissen_missing_2019}. Our approach employs the MICE (Multiple Imputation by Chained Equations) algorithm, which is handled by the \texttt{mice} package in \texttt{R} \cite{van_buuren_mice_2011}. MICE is an iterative algorithm that applies linear and logistic regression techniques in order to impute the data while conserving the relationship between the features as well as possible. The algorithm constructs \(N\) individual data sets to be modelled separately, the results of which are \emph{pooled} (combined) according to Rubin's rules \cite{rubin_multiple_1987}. In this study, 5 imputation sets were created using 20 iterations (leaving other \texttt{mice} parameters to their defaults). Because the raw features are processed, the transformation must occur either before, after, or during the imputation. To our knowledge, there are no recommended strategies for the kinds of transformations used in this study. We therefore follow the general recommendation of von Hippel of ``\emph{impute, then transform}'' \cite{von_hippel_how_2009}. As recommended by Moons et al. \cite{moons_using_2006}, the P-GRE scores are included in the imputation before preparing \texttt{ABOVE}.
\subsubsection{Machine learning analysis}\label{sec:machine_learning_analysis}
Whereas logistic regression favors interpretability (via the odds ratios), machine learning analysis (MLA) focuses on making accurate and reliable predictions. Given inputs \(x_1,\ldots,x_k\) and an output \(y\), the goal of MLA is to identify a map \(f\) such that
\begin{equation}
\label{eq:machine_learning_goal}
    y=f(x_1,\ldots,x_k)+\epsilon,
\end{equation}
where \(\epsilon\) is a prediction error. When \(y\) is categorical (e.g. binary), \(f\) is called a \emph{classifier} because it classifies a set of inputs into discrete outputs. As classifiers are seldom perfect, a major component of MLA consists of finding the optimal \(f\), i.e. minimizing \(\epsilon\). To measure "how well" a classifier is able to classify inputs we use \emph{performance metrics}. Different metrics highlight different types of behavior, meaning a classifier can score well according to one metric, but poorly according to another metric. This study employs two metrics: \emph{prediction accuracy score} and \emph{AUC-ROC score}.

The prediction accuracy score of a classifier is the portion of correctly classified cases. In terms of our data, a correctly classified case is any application for which the classifier successfully predicts whether the applicant scores above or below the cut-off score. It is typically referred to as simply the accuracy and it is often reported as a percentage. Accuracy is a number between 0\% and 100\%, where 100\% signifies a perfect classifier. While easy to interpret, accuracy is very sensitive to unbalanced output classes (see the "Domestic applicants" curve in Figure \ref{fig:PGRE_distribution} for the class imbalance faced in this study) because it does not distinguish between the output classes. For instance, if 80\% of applicants score above the cut-off, then a naive classifier predicting \emph{above} regardless of the inputs will have an accuracy score of 80\%. For this reason, accuracy should always be considered relative to class imbalance. Furthermore, because the class imbalance changes as the cut-off increases (Fig. \ref{fig:PGRE_distribution}), the interpretation of the nominal accuracy score changes. Hence, the accuracy scores of two classifiers using different cut-offs should not be compared nominally.

The AUC-ROC score is a more complex metric than accuracy. Here, ROC refers to a Receiver Operating Characteristic curve and AUC means taking the Area Under the ROC Curve. For more details regarding ROC curves, consult the supplementary material. The AUC-ROC score, or simply the AUC, is a measure of a classifier's ability to distinguish between output classes. AUC is a number between 0 and 1, where 1 signifies a perfect classifier, but a score of 0.5 is equivalent to complete guesswork. There is no universal scheme for judging AUC scores, but Hosmer et al. provides a rough guide: \(0.7\leq\text{AUC}<0.8\) is acceptable, \(0.8\leq\text{AUC}<0.9\) is excellent and \(0.9\leq\text{AUC}\) is outstanding \cite{hosmer_applied_2000}. In contrast with the accuracy score, AUC is more robust towards imbalanced output classes \cite{ling_auc_2003}, and thus AUC scores can be more reliably compared across different cut-off scores.

MLA typically consists of 2 phases: training and testing. Here, \emph{training} refers to the construction of a classifier, and \emph{testing} refers to its evaluation based on performance metrics. A typical problem in MLA known as \emph{overfitting} arises when a classifier is trained to recognize "too many details" of a data set. Thus, instead of replicating the general trend of the data set, the classifier replicates the random errors. To avoid this, it is standard practice to use different data sets for the training and testing phases by splitting the (complete) data set at random. Because random splits can have unforeseen consequences, it is common to conduct several training-testing procedures and average the performance metrics, using the standard errors of the averages as indicators for the confidence intervals. This study employs the \(K\)-fold cross-validation algorithm with \(K=10\) to prepare the random splits \cite{hastie_elements_2017}.

It is important to note that to find a perfect classifier is typically considered impossible, even if \(\epsilon=0\) for all known data. Thus, there is no correct algorithm for constructing \(f\), and in fact, there are many unique algorithms to choose from. This study employs the conditional inference forest (CIF) algorithm, which is variant of the earlier random forest algorithm \cite{breiman_random_2001,hothorn_unbiased_2006}. A random forest is comprised of an ensemble of decision trees, each of which is an independent classifier. A decision tree is an algorithmic approach to decision-making (predictions) that asks a series of yes-no questions based on the input data (e.g. "male?" and "\(\text{GPA}>3.0?\)"). The questions are determined during the training phase and are chosen to optimize performance. Each tree is given a random sample of the training set and a random selection of the input features. Predictions of the forest are then based on a majority vote among the predictions of the trees. A CIF is similar to a random forest in principle, but differs in its construction.

This study employs the CIF algorithm via the \texttt{party} package in \texttt{R} \cite{hothorn_survival_2006,strobl_bias_2007,strobl_conditional_2008}. The forests were built using 200 trees and 3 features per tree (following the recommended \(\sqrt{p}\) \cite{svetnik_random_2003}), all other parameters kept at their defaults. One of the selling points of the CIF is that it provides a natural way of measuring the importance of each feature in the model. The process of preparing the importance measures for each feature is also handled by \texttt{party}. The idea is to remove a feature from the forest and measure the resulting change in a performance measure, interpreting a larger change as the feature being more important. As described in Janitza et al., measuring AUC loss is preferred due to its robustness with imbalanced data \cite{janitza_auc-based_2013}. The importance measure is a tool for comparing the relative importance of features and should not be interpreted further \cite{auret_interpretation_2012}.

Because the importance measures focus on the impact of removing each feature separately, a backwards recursive feature elimination (RFE) procedure is conducted to study the effect of removing several features. (see e.g., \cite{hastie_elements_2017}) To restrict the scope, the procedure is only executed for P-GRE cut-offs in intervals of 30 pt. RFE is an iterative process that involves training a forest, estimating its performance, and removing the least important feature from the set of active features. Starting with all features, the process is repeated until one feature remains. The order of removal is determined by the importance measures of the forest model. The importance measures are computed using the complete model, i.e., \emph{not} during the procedure, to avoid overfitting \cite{svetnik_application_2004}. Because the importance measures vary depending on the cut-off, one would ideally prepare a removal order separately for each cut-off and conduct a unique RFE for each cut-off. However, because the importances measures are similar for different cut-offs, an average removal order is used for all cut-offs.
\subsection{A summary of the models}\label{sec:clarifying_models}
Most of the data preprocessing described in Sec. \ref{sec:on_the_data} is done for logistic regression. This includes combining racial and ethnic minorities in an underrepresented minority category; reducing the Carnegie features \emph{Carnegie Classification}, \emph{Undergraduate Population Profile}, \emph{Funding Category} and \emph{ACT selectivity category} to binary labels; combining the Barron's selectivity categories less competitive than \emph{most} and \emph{highly competitive} to a \emph{not as competitive} category; and categorizing physics programs (both undergraduate and graduate) as \emph{large} if the number of graduates is above the 75th national percentile. Because the computational difficulties of logistic regression related to multicolinearity and low-frequency categories are circumvented by the decision-tree construction of the CIF algorithm, none of these preprocessing procedures are required for the data to be compatible with the CIF models. With the exception of combining the 2017 and 2018 graduates data into minimum and maximum cases, the data is only preprocessed for the logistic regression models.

Avoiding to preprocess the data for the CIF models is actually in line with the philosophy of the predictive modelling approach. In contrast with how logistic regression emphasizes interpretability, machine learning is only interested in the relationship between the features and the response. Preprocessing the data dilutes the available information, and thus may negatively affect the predictive analysis (e.g., as in Fig. \ref{fig:PGRE_violin_by_race}).

Overall, \(19\times2\times2\) logistic regression and CIF models are studied: there are 19 unique P-GRE cut-offs under consideration, and for each cut-off, a model is constructed with and without the potential duplicate applications (Sec. \ref{sec:on_the_data}), and using the maximum and minimum number of graduates between 2017 and 2018 (Sec. \ref{sec:no_bach_phd_grads}).

\section{Results}
Because of the large volume of similar results, we will primarily present the results (odds ratios of logistic regression and feature importance measures of the CIF) for the models including the potential duplicate applications. We discuss deviations from these results where relevant.
\subsection{Key Findings}
The statistical effects from the institutional features become more involved, both in the logistic regression models and the CIF models, as the P-GRE cut-off increases. In particular, applicants from well-funded institutions with large physics programs and high research activity are more likely to score above the cut-off. The logistic regression models and the CIF models identify several examples where the effect of an institutional feature is comparable to U-GPA or gender. While U-GPA and gender are integral components of every model (as expected), the race and ethnicity of applicants did not contribute as much to our models as anticipated based on the differences in scores between racial and ethnic groups found by Miller et al. \cite{miller_typical_2019}. Overall, the logistic regression approach and the machine learning approach typically agree on whether a feature has any relevance in the model. Having said that, the odds ratios typically identify a larger set of important features that, in addition, changes as the P-GRE cut-off increases.


\begin{figure*}[t!]
    \centering
    \includegraphics[width=0.99\textwidth]{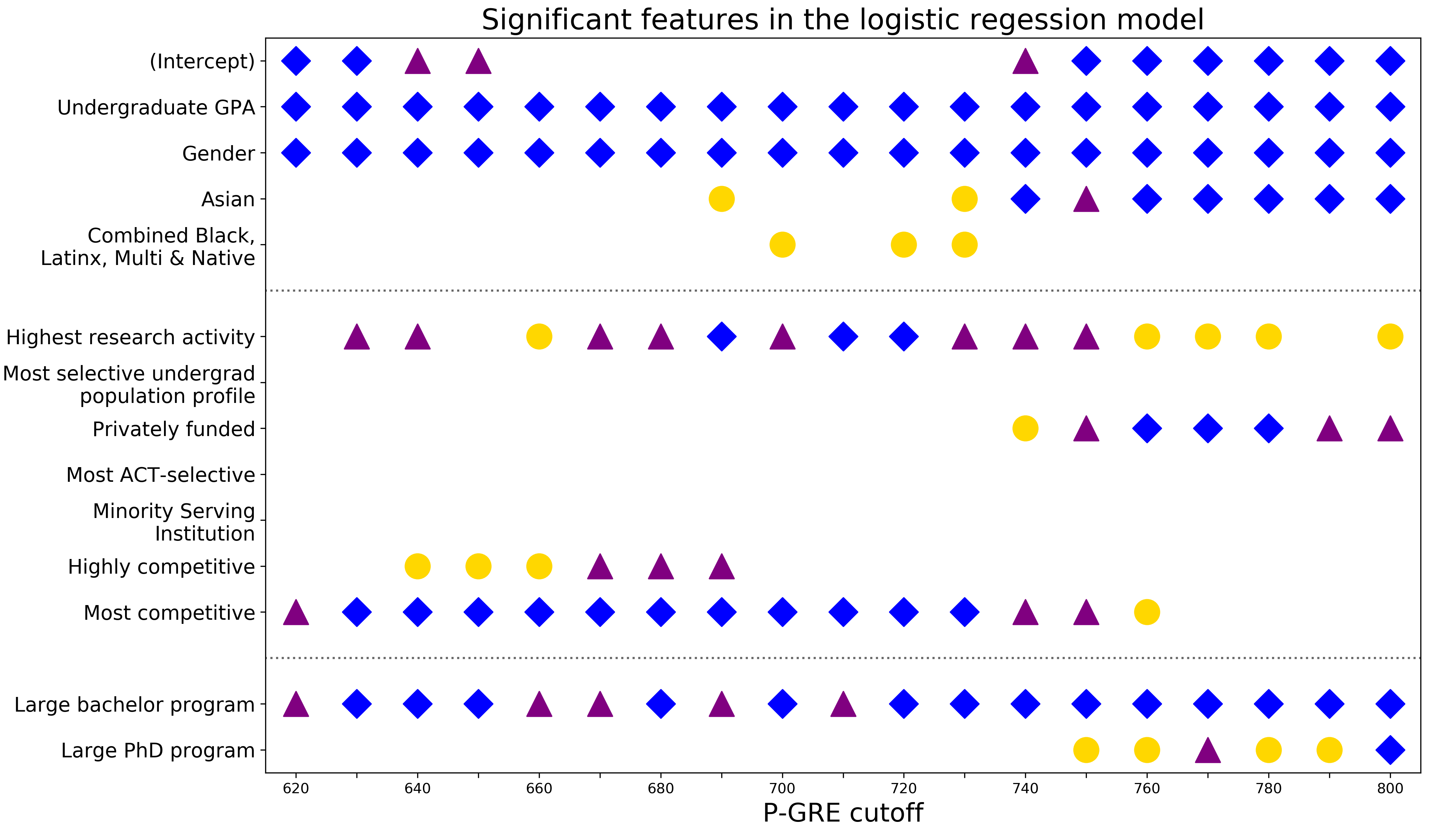}
    \caption{Summary of odds ratio significance levels of 19 independent models: Each P-GRE cut-off score (horizontal coordinate) denotes an independent maximum graduates logistic regression model that includes possible duplicates (see Sec. \ref{sec:clarifying_models}). In each model, the fields of the statistically significant features are marked with a symbol indicating the Bonferroni-corrected significance level: circle indicates \(\alpha=0.05\), triangle indicates \(\alpha=0.01\) and diamond indicates \(\alpha=0.001\). Blank fields indicate statistically insignificant results. The dotted separation lines categorize the features as applicant-related, institution-wide metrics and concerning Physics departments.}
    \label{fig:significance_diagram}
\end{figure*}

When it comes to the maximum and minimum graduates models, the contributions from the AIP features are devalued in the minimum graduates models in favor of Barron's selectivity index and high research activity. Of the maximum models, the logistic regression models favor attending an institution with a large bachelor program over a large PhD program, while the opposite is true in the CIF models. Finally, the analysis as a whole is similar for the models with and the models without possible duplicate applications. Specifically, by removing the possible duplicates, some features become less significant in the logistic regression models and the performance of some CIF models are slightly reduced.
\subsection{Logistic Regression}
\subsubsection{Significance analysis}\label{sec:Results_LR_significance_analysis}
Consider first the maximum number of graduates models. Figure \ref{fig:significance_diagram} shows a diagram indicating how the set of significant features changes between the logistic regression models as the P-GRE cut-off score is increased from 620 to 800. The features are typically significant for every or almost every cut-off, for no or few cut-offs, or for higher cut-offs. In the following we provide an overview of the applicant-related and institutional features that are statistically significant.

Of the applicant-related features, the odds ratios of U-GPA and gender are always statistically significant. However, contrary to expectations, odds ratios between applicants from different racial groups were only statistically significant for some cases. In particular, when compared to applicants identifying as white, the odds ratios for applicants identifying as Asian are significant for higher cut-offs, while the odds ratios for applicants identifying as Black, Latinx, Multi or Native are only significant for a few cut-offs in the region of \(\approx710\). This is further discussed in Sec. \ref{sec:data_processing_modelling_choices}.

When it comes to the institutional features, those statistically significant to a majority of the P-GRE cut-offs include attending a most competitive institution, an institution practicing some of the highest amounts of research activity, and an institution with a large physics bachelor program. Interestingly, attending a highly competitive institution is only significant for cut-offs between 640 and 690, while attending one of the most competitive institutions is significant for all cut-offs up to 760. Additionally, attending private universities or universities with large PhD programs becomes significant when the cut-off increases beyond \(\approx740\). In contrast, attending an MSI, a most ACT-selective institution or to graduate in a most selective undergraduate population profile is never significant, regardless of the cut-off.

In order to provide a rough overview of the difference between the maximum and minimum number of graduates models, Table \ref{tab:significance_count} shows the fraction of P-GRE cut-offs for which each feature is significant. Note that the table also separates models with and without the possible duplicate applications. By removing the possible duplicates, the general significance of the features decreases. The change does not seem to originate in any particular feature as, with the exception of U-GPA, gender and most competitive, the fraction of significant cut-offs is reduced for all features. Compared to the maximum graduates models, the typical significance of attending a large bachelor program is considerably lower in the minimum graduates models. Notably, the difference corresponds with an improvement in the fraction of significant cut-offs for attending a competitive school or an institution with high research activity, thus suggesting the variables may suffer from a confounding issue (see Sec. \ref{sec:limitations} for a discussion).

\begin{table}
    \centering
    \caption{Fraction of P-GRE cut-offs for which the odds ratio of each feature is statistically significant in the logistic regression models. There are 19 logistic regression models in each category (see Sec. \ref{sec:clarifying_models}). The first column (the maximum graduates models with possible duplicates) corresponds to the significance diagram (Fig. \ref{fig:significance_diagram}).}
    \label{tab:significance_count}
    \begin{tabular}{lcccc}
        \toprule
        \multicolumn{1}{c}{\multirow{2}{*}{Variable}} &
        \multicolumn{2}{c}{Maximum} & \multicolumn{2}{c}{Minimum} \\
        \cmidrule(lr){2-3}\cmidrule{4-5}
        & With & Without & With & Without \\
        \midrule
        (Intercept)                    & 0.63 & 0.58 & 0.58 & 0.58 \\
        Undergraduate GPA              & 1.00 & 1.00 & 1.00 & 1.00 \\
        Gender                         & 1.00 & 1.00 & 1.00 & 1.00 \\
        Combined B, L, M \& N          & 0.26 & 0.00 & 0.21 & 0.00 \\
        Asian                          & 0.53 & 0.37 & 0.47 & 0.42 \\
        Highly competitive             & 0.26 & 0.26 & 0.63 & 0.26 \\
        Most competitive               & 0.79 & 0.89 & 1.00 & 1.00 \\
        Highest research activity      & 0.89 & 0.63 & 1.00 & 0.63 \\
        Most selective UG p. profile   & 0.00 & 0.00 & 0.00 & 0.00 \\
        Most ACT selective             & 0.00 & 0.00 & 0.00 & 0.00 \\
        Privately Funded               & 0.42 & 0.21 & 0.32 & 0.21 \\
        Minority Serving Institution   & 0.00 & 0.00 & 0.00 & 0.00 \\
        Large bachelor program         & 1.00 & 0.74 & 0.37 & 0.21 \\
        Large PhD program              & 0.32 & 0.05 & 0.32 & 0.11 \\
        \bottomrule
    \end{tabular}
\end{table}


Considerable changes in the set of significant features are only observed for large changes in the cut-off score. We therefore only discuss the odds ratios corresponding to cut-offs 650, 710 and 770, representing the lower, middle and higher regions, respectively.
\subsubsection{Odds ratios}\label{sec:Results_LR_odds_ratios}
The odds ratios for the maximum and minimum number of graduates models are shown in Tables \ref{tab:odds_ratios} \subref{tab:odds_ratios_best_case} and \subref{tab:odds_ratios_worst_case} respectively. First and foremost, improving one's undergraduate GPA by one standard deviation, roughly equivalent of improving a B to a B+, improves the odds of scoring above the cut-off by at minimum a factor of 2.5 (increases to \(\approx2.8\) for higher cut-offs). This substantial increase in odds reflects the importance of U-GPA in admissions expressed by both admission committees and prospective students \cite{potvin_investigating_2017,chari_understanding_2019}. Additionally, the odds of scoring above the cut-off is \(1/0.17\approx5.9\) times greater for male applicants than for female applicants. The odds ratios of U-GPA and gender are consistent for all P-GRE cut-offs in both the maximum and minimum number of graduates models.

While the benefit of attending a competitive institution diminishes as the P-GRE cut-off increases from 650 to 710 and 770, attending one of the most competitive institutions is always preferable to a highly competitive institution. For cut-offs 650 and 710, the odds-increase from attending a most competitive school is similar to the applicant increasing their U-GPA from a B to a B+. The model also finds institutional funding and high levels of research activity to be important factors. For high P-GRE cut-offs (e.g. 770), the odds of scoring above the cut-off is about 2 times as large for applicants who attended a private university compared to applicants who attended a public university. Similarly, for applicants attending a university that practices some of the highest levels of research activity, the odds ratio is roughly 1.6-2.0 depending on the cut-off.

\begin{table*}[t!]
    \centering
    \caption{Odds ratios for P-GRE cut-off scores 650, 710 and 770 of the logistic regression models with possible duplicates (see Sec. \ref{sec:clarifying_models}). The maximum and minimum graduates models are separated in Tables \protect\subref{tab:odds_ratios_best_case} and \protect\subref{tab:odds_ratios_worst_case} respectively. Statistically significant odds ratios are marked with asterisks (see below \protect\subref{tab:odds_ratios_worst_case}). Note that \(\tilde{p}=14p\) refers to the Bonferroni-corrected \(p\)-values.}
    \label{tab:odds_ratios}

\subfloat[Maximum bachelor/PhD graduates models\label{tab:odds_ratios_best_case}]{
\begin{tabular}{lrrrrrrrrrrrr}
    \toprule
        \multicolumn{1}{c}{\multirow{2}{*}{Variable}} &
        \multicolumn{3}{c}{cut-off 650} &&
        \multicolumn{3}{c}{cut-off 710} &&
        \multicolumn{3}{c}{cut-off 770} &  \\
    \cmidrule{2-4}\cmidrule{6-8}\cmidrule{10-12}
        & \multicolumn{1}{c}{OR} & \multicolumn{2}{c}{95\% CI} &
        & \multicolumn{1}{c}{OR} & \multicolumn{2}{c}{95\% CI} &
        & \multicolumn{1}{c}{OR} & \multicolumn{2}{c}{95\% CI} & \\
    \midrule
        (Intercept)                             & 1.72 & [1.09, & 2.72] & \taast  & 0.81 & [0.53, & 1.25] &         & 0.29 & [0.18, & 0.47] & \taaast \\
        Undergraduate GPA                       & 2.53 & [2.12, & 3.03] & \taaast & 2.63 & [2.22, & 3.11] & \taaast & 2.87 & [2.40, & 3.42] & \taaast \\
        Gender                                  & 0.16 & [0.10, & 0.24] & \taaast & 0.18 & [0.12, & 0.27] & \taaast & 0.17 & [0.11, & 0.25] & \taaast \\
        Asian                                   & 1.37 & [0.73, & 2.54] &         & 1.61 & [0.95, & 2.71] &         & 2.20 & [1.34, & 3.59] & \taaast \\
        Combined B, L, M \& N                   & 0.68 & [0.41, & 1.11] &         & 0.65 & [0.42, & 1.00] &         & 0.76 & [0.49, & 1.19] &         \\
        Highly competitive                      & 1.98 & [1.10, & 3.58] & \tast   & 1.55 & [0.95, & 2.52] &         & 1.22 & [0.76, & 1.93] &         \\
        Most competitive                        & 3.40 & [1.76, & 6.56] & \taaast & 2.79 & [1.53, & 5.07] & \taaast & 1.62 & [0.97, & 2.73] &         \\
        Doc. inst. w/ highest research activity & 1.68 & [1.00, & 2.81] & \tast   & 1.94 & [1.23, & 3.04] & \taaast & 1.63 & [1.06, & 2.51] & \tast   \\
        Most selective UG population profile    & 1.72 & [0.68, & 4.37] &         & 1.85 & [0.73, & 4.68] &         & 1.75 & [0.71, & 4.32] &         \\
        Most ACT selective                      & 0.66 & [0.26, & 1.69] &         & 0.49 & [0.19, & 1.27] &         & 0.66 & [0.25, & 1.75] &         \\
        Privately funded                        & 0.96 & [0.57, & 1.62] &         & 1.42 & [0.91, & 2.21] &         & 2.07 & [1.38, & 3.11] & \taaast \\
        Minority Serving Institution            & 1.17 & [0.62, & 2.21] &         & 1.05 & [0.60, & 1.82] &         & 1.13 & [0.64, & 1.99] &         \\
        Large bachelor program                  & 1.96 & [1.23, & 3.12] & \taaast & 1.74 & [1.15, & 2.65] & \taast  & 1.94 & [1.28, & 2.92] & \taaast \\
        Large PhD program                       & 1.28 & [0.72, & 2.25] &         & 1.31 & [0.83, & 2.08] &         & 1.62 & [1.07, & 2.45] & \taast  \\
    \bottomrule
\end{tabular}}
\hfill
\subfloat[Minimum bachelor/PhD graduates models\label{tab:odds_ratios_worst_case}]{
\begin{tabular}{lrrrrrrrrrrrr}
    \toprule
        \multicolumn{1}{c}{\multirow{2}{*}{Variable}} &
        \multicolumn{3}{c}{cut-off 650} &&
        \multicolumn{3}{c}{cut-off 710} &&
        \multicolumn{3}{c}{cut-off 770} &  \\
    \cmidrule{2-4}\cmidrule{6-8}\cmidrule{10-12}
        & \multicolumn{1}{c}{OR} & \multicolumn{2}{c}{95\% CI} &
        & \multicolumn{1}{c}{OR} & \multicolumn{2}{c}{95\% CI} &
        & \multicolumn{1}{c}{OR} & \multicolumn{2}{c}{95\% CI} & \\
    \midrule
        (Intercept)                             & 2.03 & [1.31, & 3.15] & \taaast & 0.94 & [0.61, & 1.45] &         & 0.34 & [0.22, & 0.53] & \taaast \\
        Undergraduate GPA                       & 2.50 & [2.08, & 2.99] & \taaast & 2.59 & [2.19, & 3.08] & \taaast & 2.82 & [2.35, & 3.39] & \taaast \\
        Gender                                  & 0.17 & [0.11, & 0.25] & \taaast & 0.19 & [0.13, & 0.28] & \taaast & 0.17 & [0.12, & 0.26] & \taaast \\
        Asian                                   & 1.41 & [0.72, & 2.75] &         & 1.63 & [0.96, & 2.77] &         & 2.19 & [1.35, & 3.56] & \taaast \\
        Combined B, L, M \& N                   & 0.66 & [0.41, & 1.07] &         & 0.65 & [0.42, & 0.99] & \tast   & 0.77 & [0.50, & 1.17] &         \\
        Highly competitive                      & 2.21 & [1.25, & 3.93] & \taaast & 1.68 & [1.00, & 2.80] & \tast   & 1.35 & [0.83, & 2.18] &         \\
        Most competitive                        & 3.69 & [2.07, & 6.55] & \taaast & 3.09 & [1.89, & 5.05] & \taaast & 1.87 & [1.17, & 3.00] & \taast  \\
        Doc. inst. w/ highest research activity & 1.72 & [1.02, & 2.92] & \tast   & 2.10 & [1.33, & 3.33] & \taaast & 1.83 & [1.20, & 2.79] & \taaast \\
        Most selective UG population profile    & 1.56 & [0.60, & 4.05] &         & 1.72 & [0.62, & 4.76] &         & 1.59 & [0.62, & 4.07] &         \\
        Most ACT selective                      & 0.75 & [0.28, & 2.03] &         & 0.56 & [0.19, & 1.67] &         & 0.74 & [0.28, & 1.99] &         \\
        Privately funded                        & 0.88 & [0.53, & 1.45] &         & 1.27 & [0.81, & 1.98] &         & 1.88 & [1.27, & 2.79] & \taaast \\
        Minority Serving Institution            & 1.11 & [0.60, & 2.06] &         & 1.00 & [0.58, & 1.74] &         & 1.11 & [0.64, & 1.93] &         \\
        Large bachelor program                  & 1.44 & [0.91, & 2.29] &         & 1.28 & [0.84, & 1.94] &         & 1.47 & [0.99, & 2.20] &         \\
        Large PhD program                       & 1.43 & [0.83, & 2.45] &         & 1.34 & [0.86, & 2.08] &         & 1.54 & [1.05, & 2.26] & \tast   \\
    \bottomrule
    & \multicolumn{4}{l}{\(\text{\taaast}:\tilde{p}\leq0.001\)}
    & \multicolumn{4}{l}{\(\text{\taast}:\tilde{p}\leq0.01\)}
    & \multicolumn{4}{l}{\(\text{\tast}:\tilde{p}\leq0.05\)}
\end{tabular}}
\end{table*}

Applicants from institutions with large physics programs typically also score higher. In the maximum graduates number of models, having attended a university with one of the largest undergraduate physics programs improves the odds of scoring above the P-GRE cut-off by a factor of about 1.7-2.0 (typically closer to 2.0). When the cut-off is high, a similar effect is seen for students attending a university offering a large graduate program (an odds ratio of about 1.6). In the minimum number of graduates models, the odds ratios are only statistically significant for the highest cut-offs (\(\geq760\)). They are also typically smaller than the corresponding odds ratios in the maximum graduates models. The only statistically significant example in Table \ref{tab:odds_ratios_worst_case} is the odds ratios for attending an institution with one of the largest PhD programs.

The remaining variables, i.e., most ACT-selective, most selective undergraduate population profile and MSI, contribute little to none.

\subsection{Conditional Inference Forest}\label{sec:Results_CIF}
\begin{figure*}[p!]
    \centering
    \includegraphics[width=\textwidth]{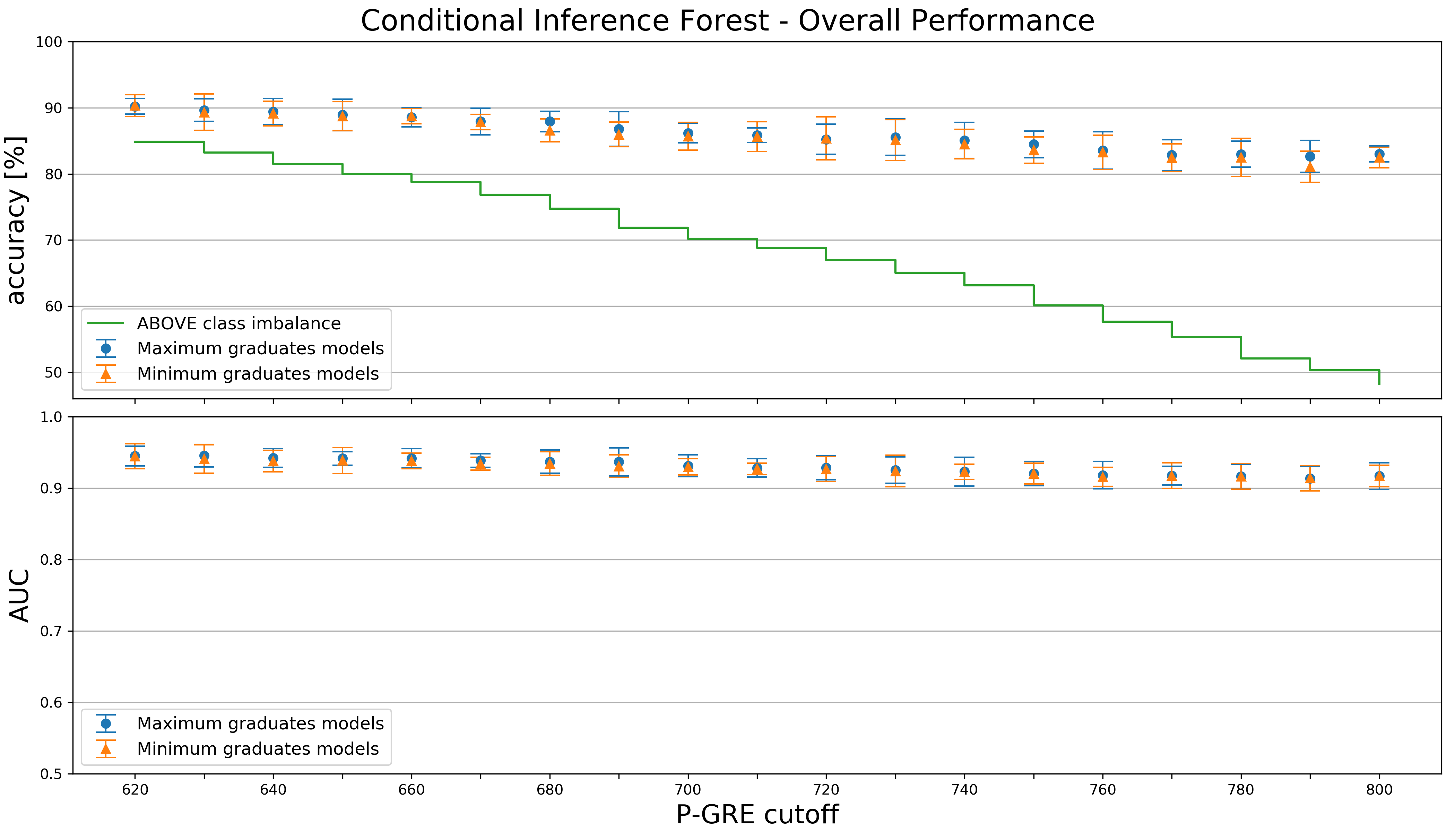}
    \caption{Overall performance of the conditional inference forest. The standard errors of the \(K\)-fold (\(K=10\)) estimates are indicated by the error bars. While the \texttt{ABOVE} class imbalance is very high for lower cut-offs, the accuracy standard errors are always above the imbalance level. The AUC score is mostly above 0.9, which Hosmer et al. categorizes as ``outstanding'' \cite{hosmer_applied_2000}.}
    \label{fig:CIF_overall_performance}
\end{figure*}

The general performance of the CIF models is shown in Figure \ref{fig:CIF_overall_performance}. Alongside the accuracy score is the class imbalance, which provides the baseline from which the accuracy score is interpreted. Because the imbalance is considerably high for lower cut-offs, the accuracy score is more representative of the CIF's ability to identify applicants scoring above the cut-off when the cut-off is higher (as the imbalance decreases as the cut-off score increases, the accuracy becomes increasingly more representative). However, because the imbalance level is outside the standard errors of the \(K\)-fold estimate, it is reasonable to conclude that the CIF is not simply predicting the majority class. Additionally, the AUC score is mostly \emph{outstanding} (>0.9) and, more importantly, very stable with respect to changes in the P-GRE cut-off. The stability of the AUC coupled with the high score suggests that the results of the model may be reasonably interpreted, that is, that the feature importances provide a reasonable picture of the relationship between the features and the output for all P-GRE cut-offs. Because of the similarity in performance between the maximum and minimum graduates CIF models, we present only the maximum graduates models going forward.

\begin{figure*}[p!]
    \centering
    \includegraphics[width=\textwidth]{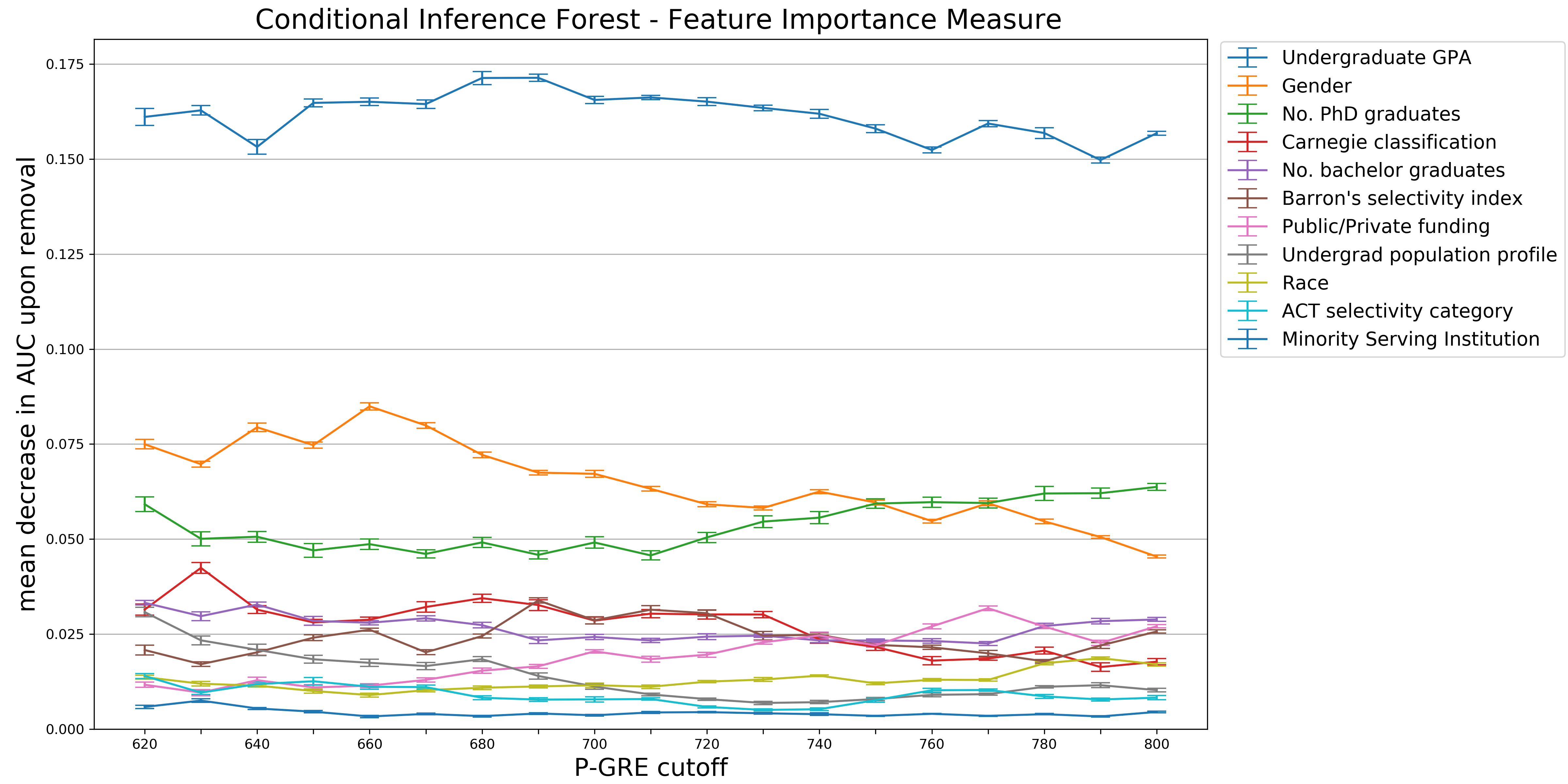}
    \caption{Conditional inference forest feature importance measures. The standard errors of the \(K\)-fold (\(K=10\)) estimates are indicated by the error bars. Features included in the CIF are listed in the legend in decreasing order of average (across all cut-offs) decrease in AUC upon removal.}
    \label{fig:CIF_feature_importance_measures}
\end{figure*}

Figure \ref{fig:CIF_feature_importance_measures} graphs the change in the importance measure of the features as the cut-off increases. The plot shows evidence of distinct groups of features with similar importances. The first group consists only of undergraduate GPA, whose importance measure is about 2 times higher than any other feature. The next group consists of gender and no. PhD graduates, which stand out when compared to the remaining group of the least important features (see Sec. \ref{sec:machine_learning_analysis} for how to interpret the importance measure). With the exception of some minor variation, the importance measure of U-GPA is fairly stable across all models. Notably however, while the importance measure decreases for gender as the cut-off increases, it simultaneously increases for no. PhD graduates. Hence, for cut-offs greater than \(\approx750\), the model finds a greater statistical difference between applicants scoring above and below the cut-off when given the no. PhD graduates compared to an applicant's gender. Proportional to their own importance measures, several features in the remaining group undergo large changes in importance measures. However, because these variations are small when compared to U-GPA, gender and the no. PhD graduates, they should not be overemphasized.

\begin{figure*}[p!]
    \centering
    \includegraphics[width=\textwidth]{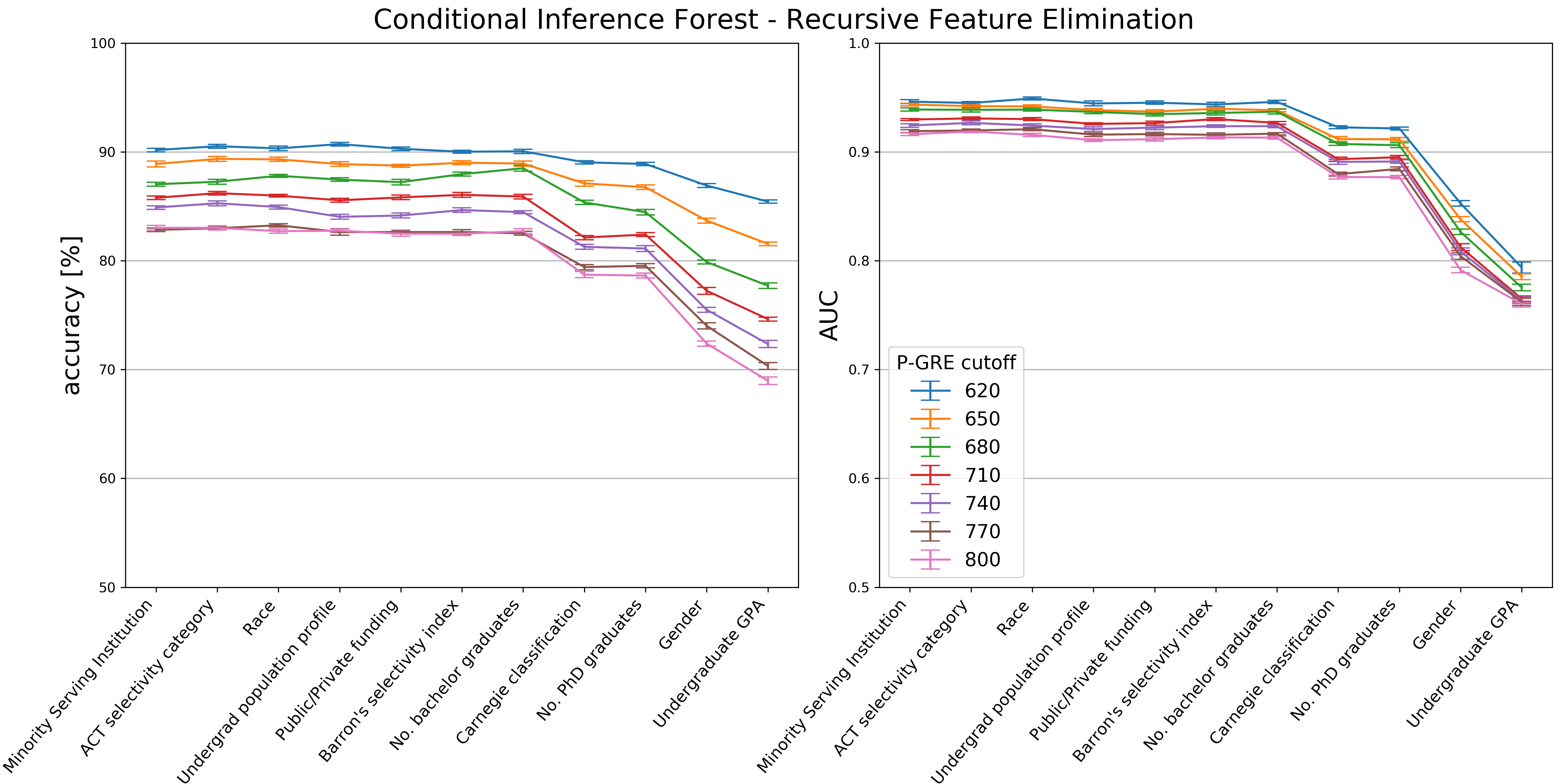}
    \caption{Conditional inference forest feature elimination procedure. The standard errors of the \(K\)-fold (\(K=10\)) estimates are indicated by the error bars. The features are eliminated from left to right, where the named feature is currently the least important feature, and thus the next to be dropped from the model.}
    \label{fig:CIF_feature_elimination}
\end{figure*}

\begin{figure*}[p!]
    \centering
    \includegraphics[width=\textwidth]{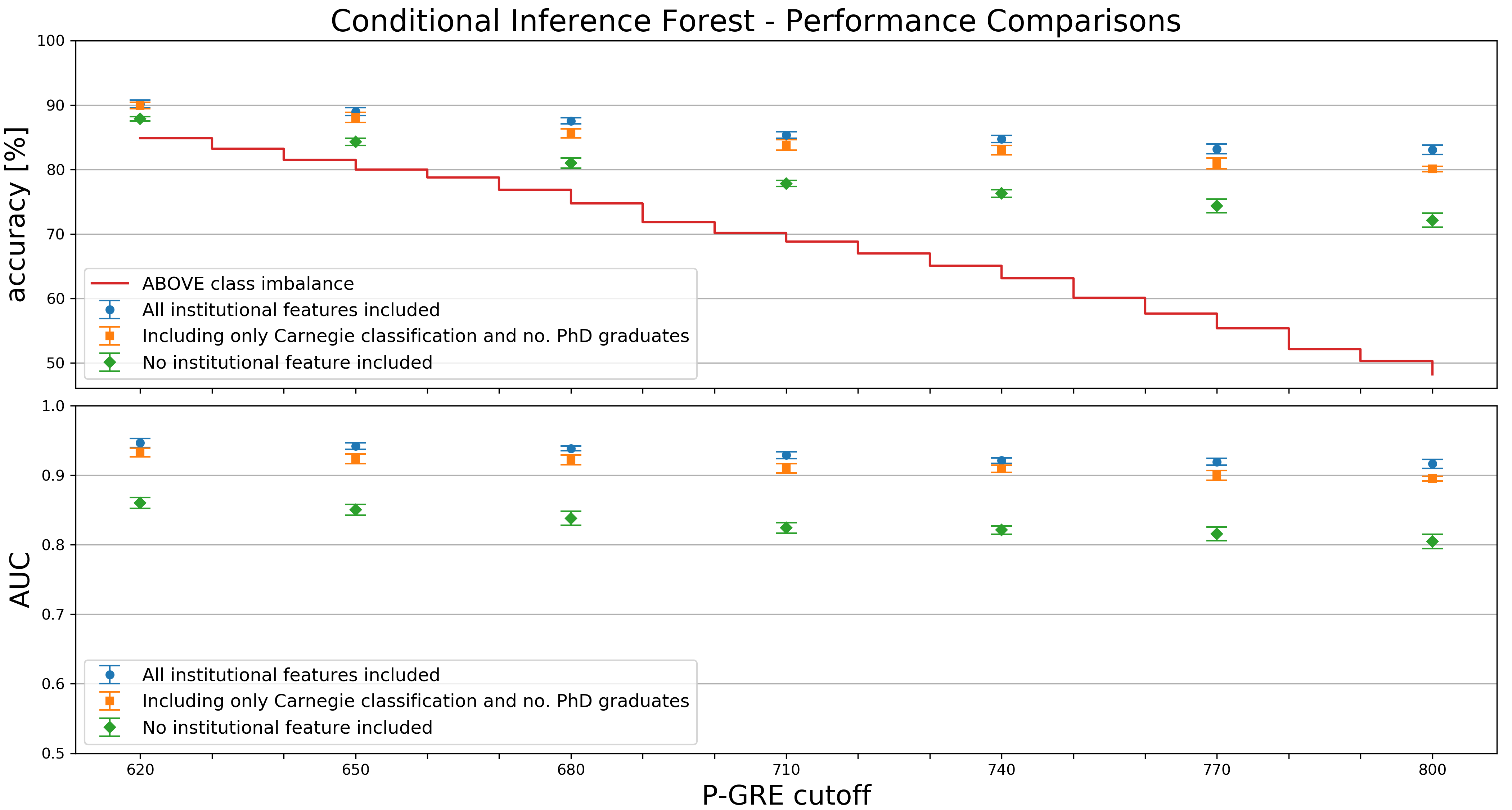}
    \caption{Performance comparison between conditional inference forests with all features, with 2 institutional features, and without institutional features (only U-GPA, gender and race). The standard errors of the \(K\)-fold (\(K=10\)) estimates are indicated by the error bars. The significant improvement in performance by the simple addition of 2 institutional features suggests that the contribution from the institutional features is captured by a few features.}
    \label{fig:CIF_performance_comparison}
\end{figure*}

The results of the feature elimination procedure are shown in Figure \ref{fig:CIF_feature_elimination}. The diagram is arranged such that the features are removed left to right, starting from a complete model and ending with a model that only includes U-GPA (i.e., the named feature at a given horizontal coordinate currently has the lowest importance measure). The accuracy and AUC scores largely agree on the effect of removing a feature. Specifically, removing MSI through Barron's selectivity index has no detrimental effect to the CIF's accuracy and AUC, and despite the high importance measure of the Carnegie classification, the model does not perform worse once it is removed either. Using only the three features with the highest importance measures (U-GPA, gender and number of PhD graduates), the CIF is able to score \(\approx0.9\) on the AUC metric and roughly between 80\% and 90\% on the accuracy metric.

Figure \ref{fig:significance_diagram} shows that the set of statistically significant features in the logistic regression models changes as the P-GRE cut-off score increases (e.g. whether the institution is privately funded is only significant for cut-offs \(\geq740\)). A similar change is not present in the importance measures of the CIF models (Fig. \ref{fig:CIF_feature_importance_measures}), which, in contrast with the odds ratios, preserve the feature groups described above. In particular, the features: U-GPA, gender and number of PhD graduates, are the three most important features for every cut-off score. Because the importance measures of the remaining features are consistently lower by a considerable margin for all cut-off scores, the set of important features in the CIF models is very robust towards changes in the cut-off score.

As a final check for whether the added performance can be attributed to including the institutional features, the performance of the full CIF is compared to a CIF excluding all institutional features, and a CIF including the number of PhD graduates and the Carnegie classification. The results of the comparison is summarized in Figure \ref{fig:CIF_performance_comparison}: The addition of only two institutional features makes a considerable improvement for both metrics, regardless of the cut-off. Hence, the added performance is reasonably attributed to the inclusion of institutional features.


\section{Discussion}
\subsection{Research Questions}
This study investigated four research questions (RQs) that we address in order.
\begin{enumerate}
    \item To what extent does the applicant's undergraduate institution influence whether they are able to attain a minimum P-GRE score expected by an admissions committee?
    \item To what degree do the institutional effects compare to known effects such as U-GPA, gender and race?
    \item How do the results depend on the specific cut-off chosen by the admissions office?
    \item How well do the conventional and machine learning approaches agree on RQs 1, 2 and 3?
\end{enumerate}

Regarding RQ 1, the institutional background helps explain whether a student scores above a given P-GRE cut-off. Consider a cut-off score of 710, which is just above the most common cut-off score of 700. In the logistic regression models (see Table \ref{tab:odds_ratios}), applicants from competitive institutions with large physics programs, practicing high levels of research are statistically more likely to score above the cut-off than other applicants. Similarly, the size of physics programs (number of graduates) and the institution-wide Carnegie classification are integral components of the predictive capacity of the CIF models (see Sec. \ref{fig:CIF_feature_elimination}). Hence, the models suggest that to employ a cut-off score of 710 not only limits access to racial and ethnic minorities \cite{miller_typical_2019}, but also to applicants from smaller universities with less resources that are less competitive and practice lower (not necessarily among the lowest) levels of research. Similar observations are found for every other cut-off in the CIF models. In the case of the logistic regression models, the set of statistically significant institutional features varies depending on the cut-off, but the overall interpretation is similar: To include institutional data in the analysis certainly helps explain whether a student scores above the cut-off, regardless of the chosen cut-off.

Now, is it necessary to include a complete description of an applicant's undergraduate background? Figure \ref{fig:CIF_feature_elimination} suggests that this is probably not the case as a large portion of the institutional data does not contribute to the models. Moreover, because the performance of the CIF does not decrease as the Carnegie classification is removed, there is also reason to suspect that the institutional features may share information. The independence of the features is discussed more in detail in Sec. \ref{sec:limitations}.

The modelling and machine learning approaches disagree somewhat with respect to RQ 2. In the logistic regression model, the odds ratios for U-GPA is comparable to admission competitiveness (roughly 2-3), while the odds ratios for gender is just shy of 6.0. In contrast, U-GPA is by far the most important feature in all CIF models. Meanwhile, the feature importance measure of gender is similar to the number of PhD graduates, particularly for higher cut-offs (\(\geq750\)). Because neither approach placed as much emphasis on race and ethnicity, it is unreasonable to judge the overall effect of institutional data by comparing it to the effects of race and ethnicity in the models. Despite disagreeing on some of the finer details, both approaches find examples where the effects from institutional data, e.g. admission competitiveness and the size of Physics departments, are comparable to U-GPA and gender. The most clear-cut example is shown in figure \ref{fig:CIF_performance_comparison}, which demonstrates that to replace a CIF model without institutional features with a similar CIF model that includes the Carnegie classification and number of PhD graduates provides a blanket improvement in the accuracy and AUC scores for every P-GRE cut-off.

Finally, we address RQs 3 and 4 together. First and foremost, both approaches have identified statistically significant differences in P-GRE scores of applicants with different institutional backgrounds. Having said that, the specifics regarding the statistical difference and the extent to which it is explained by different institutional backgrounds depends on the model and cut-off in question. For instance, the significance level of odds ratios vary to such an extent that some features are only relevant for a select few cut-offs (e.g. private/public institution for higher cut-offs). The importance measures of the CIF models are much more stable across cut-offs, but lacks the interpretability of the odds ratios. Nevertheless, while the set of useful features changes with the cut-off, institutional features always contribute to the analysis. Here, logistic regression disagree with the CIF on the set of useful features and their importance to the model, but both recognize useful institutional features for every cut-off score.
\subsection{Limitations}\label{sec:limitations}
Central to this study is the question of whether the institutional background of an applicant can be reliably measured, or estimated, with the available data. Here, "institutional background" is used in an extended sense that includes the applicant's experiences in relation to attending a particular institution. Our data certainly does not allow for quantifying the effects of such experiences as studying in an encouraging environment or at an institution with a large array of opportunities. However, data such as the Carnegie features and the number of graduating bachelor and PhD students likely capture some aggregate effect of studying at different types of institutions. In addition, these features were found to be important in our models, suggesting that there is a statistical difference between the applicants that is dependent on the institutions.

Because the universities considered in this study are typically highly regarded, the data likely suffers from a selection bias effect, favoring prospective students with higher grades and GRE scores. In a 2018 survey of prospective students from racial and ethnic minorities, Cochran et al. identified concerns regarding GRE scores and undergraduate GPA as commonly expressed barriers to apply to physics graduate programs \cite{cochran_identifying_2018}. Indeed, this is reflected in the P-GRE distribution of the applicants in our data set: Figure \ref{fig:PGRE_distribution} shows that the applicants consistently score as high or higher than the national averages, thus implying our data set consists of a biased selection of all prospective students (the data set comprises an upper limit of \(\approx18\%\) of all P-GRE test-takers in 2017-18 \cite{educational_testing_service_gre_2019}. Because of this selection bias, the distributions of the other features in our data set are likely also biased. Most prominently, the selection bias will disproportionately affect women, and racial and ethnic minorities \cite{porter_women_2019,ivie_beyond_2018}. The problem of selection bias and its consequences for Physics education research as a whole was recently discussed in Kanim and Cid \cite{kanim_demographics_2020}. Our findings should thus be considered in light of our biased sample and their discussion.

A related, but different issue is that applicants are more likely to have attended large programs by virtue of there being more prospective students from larger programs than smaller programs. This can be seen in our data from the median number of Bachelor graduates. Whereas the national median was 8 in both 2017 and 2018 \cite{nicholson_roster_2018,nicholson_roster_2019}, the median in our data is 27 (2017) and 30 (2018), i.e., more than 3 times as many. Consequently, our data consists of a larger fraction of applicants from larger programs than usual, and thus the distributions of all the features in our data are likely primarily determined by applicants from larger programs. This also contributes to the selection bias discussed above.

Another methodological problem is the question of whether the different institutional variables attempt to describe the same effect, implying a possible problem of correlation, or even multicollinearity, between the features. The number of Bachelor and PhD graduates are particularly sensitive to this issue as they both represent a measure of the size of physics departments. Indeed, the features share a positive correlation of roughly 0.7. Both approaches present evidence in favor of there being some degree of relationship between the features. For instance, when comparing the minimum and maximum graduates logistic regression models, Table \ref{tab:significance_count} shows that the difference in the fraction of P-GRE cut-offs for which the size of bachelor and PhD programs are significant is similar to the same difference for attending a competitive school or an institution with high research activity. As it is not uncommon for institutions with larger programs to be more competitive or practice higher levels of research, we suspect that some statistical relationship between these features is likely. A more direct example is seen in Figure \ref{fig:CIF_feature_elimination}, where the removal of the Carnegie classification during the feature elimination procedure does not deteriorate the performance by any measurable amount. This indifference suggests that the information contained in the Carnegie classification, which is known to be considerable due to Carnegie's high importance measure (see Figure \ref{fig:CIF_feature_importance_measures}), is also contained within the remaining set of features (U-GPA, gender and number of PhD graduates). As a final example, the performance comparison (Figure \ref{fig:CIF_performance_comparison}) shows that most of the overall effect of the institutional influence can be described by a limited selection of institutional features.
\subsection{Data processing and modelling choices}\label{sec:data_processing_modelling_choices}
A major difficulty for the logistic regression approach is the need for data processing, especially in the context of losing information by unfortunate modelling choices. The most prominent example in this study is the combination of racial and ethnic groups into a single underrepresented minority group. As suggested by Figure \ref{fig:PGRE_violin_by_race}, the lack of race features being important in the logistic regression model may actually be a case of Simpson's paradox (information loss due to combining data \cite{simpson_interpretation_1951}). That is, because the combined P-GRE distribution of URM applicants resembles the P-GRE distribution of white applicants (see Figure \ref{fig:PGRE_violin_by_race}), and because the race feature was one-hot encoded using ``white'' as reference level, the difference between the distributions is not large enough to be statistically significant. In comparison, the distribution is much more skewed for Asian applicants, and thus the difference becomes statistically significant for higher cut-offs. Other examples include the Carnegie classification and undergraduate population profile, which were essentially reduced from multi-level categorizations to simple binaries. Estimating the amount of meaningful information lost for these features is particularly complicated because of the high number of low-frequency categories.

Compared to the logistic regression approach, the CIF avoids the data processing issues described above. When processing categorical features for inferential modelling, the features must remain interpretable. However, because the CIF does not require the combination of categorical levels to be meaningful, a tree node can find the optimal grouping of categories without regard to interpretation. Indeed, the construction of the CIF algorithm allows it to naturally handle unprocessed data without suffering the same issues as logistic regression (and other machine learning methods that require preprocessing the data). As a result, the CIF is able to identify statistical properties much more easily than logistic regression. An example of this effect is seen in Barron's selectivity index: Whereas the odds ratios decrease and become less significant as the P-GRE cut-off increases (Table \ref{tab:odds_ratios}), the feature importance is relatively stable with respect to changes in the cut-off (Figure \ref{fig:CIF_feature_importance_measures}). 

Furthermore, compared to the odds ratios of logistic regression, the importance measures of the CIF are more effective and provide a clearer picture. The framework of logistic regression assumes that every feature is a distinct component of the response (eq. \eqref{eq:odds_equation}). In contrast, a tree in the CIF will only include a feature if its found to be important enough (see Sec. \ref{sec:machine_learning_analysis}). Hence, if a particular feature is always less important than the other features in every tree (recall each tree is built on a subset of the features), then its importance measure will be 0. A similar mechanism is not present in the logistic regression framework, which will always try to interpret every feature as an integral component of the model. Accordingly, the importance measures more accurately reflect the degree to which the features are associated with the response. Indeed, note that the set of features essential to the model is always larger in the logistic regression models, and in addition, changes as the P-GRE cut-off increases. For example, the odds ratio for attending a privately funded institution is only statistically significant for cut-off scores \(\geq740\) (Figure \ref{fig:significance_diagram}). By relaxing the necessary assumptions of the logistic regression framework, we get a more effective tool for identifying the relationship between the features and the response, albeit one that is harder to interpret.

The effects of unfortunate modelling choices in logistic regression models depends, in the end, on the data. In our case, the combining of racial and ethnic minorities in an underrepresented minority category has likely influenced how racial and ethnic information is treated model. Similarly, the significance of other processed features may also have been diminished. That being said, we have conducted two very different analyses (inferential vs. predictive modelling) and found similar results. It is therefore unlikely that the choices unique to each approach have affected the overall results of the analysis.
\subsection{Future work}
The present study has looked into how the undergraduate institutions of applicants may influence the physics graduate admissions process by studying its statistical relationship with P-GRE cut-off scores. Lacking from this analysis is an understanding of whether institutional influence may exert its primary effect at a different stage in the admissions process. For example, it is known that a number of bachelor students that are interested in further studies eventually decide not to apply \cite{cochran_identifying_2018}. While some cases arise due to personal or financial concerns, some students may not have received the preparation or encouragement necessary for motivating further studies. If such motivation plays a significant role for students unsure of whether to pursue a career in physics, then one would expect that prospective students from institutions with PhD programs would be more likely to apply to graduate programs. Additionally, it is worth considering whether these prospective students are more likely to apply to any graduate program in general, or simply the program at their undergraduate institution.

\section{Conclusion}
    The present work has studied the effects of institutional influence on graduate program admissions by modelling a hard physics GRE cut-off score with application data from five Midwestern universities. For completeness, all possible cut-off scores between 620 and 800 (32nd and 67th percentile) have been analyzed, although most admissions employ a cut-off of 700. The analysis has been conducted using both inferential and predictive modelling based on logistic regression and the conditional inference forest algorithm respectively. Both approaches identify the known effects of undergraduate GPA and gender, but do not emphasize a statistical difference between applicants from different racial and ethnic minorities as expected from earlier work \cite{miller_typical_2019}. However, this apparent contradiction with past work can likely be understood as a combination of a Simpson's paradox and selection bias among the applicants. Both approaches identified cases where the impact of institutional features were comparable to the known effects of undergraduate GPA and gender. Overall, the two approaches agree on the analysis as a whole, but disagree on the result of increasing the P-GRE cut-off. In terms of the odds ratios, increasing the cut-off places more significance on institutional features associated with competitive schools, private funding, large physics programs and high research activity. On the other hand, the added performance when including institutional features can be attributed to a small number of features.

    In conclusion, when analyzing graduate program applications we recommend including information regarding the applicants' bachelor institutions. Moreover, due to the innate flexibility and precision of the conditional inference forest algorithm, combined with the large variety of data structures seen in application data, we also recommend the forest algorithm as well as the predictive analysis approach in general. Based on these findings and its known problems of limiting underrepresented racial and ethnic minorities, we advocate against the practice of using GRE cut-off scores in admissions.
    
    \hspace{1cm}
    
\begin{acknowledgments}
    This project was supported by the Michigan State University College of Natural Sciences, the Lappan-Phillips Foundation, and the Norwegian Agency for Quality Assurance in Education (NOKUT), which supports the Center for Computing in Science Education. This project has also received support from the INTPART project of the Research Council of Norway (Grant No. 288125) and the Thon foundation.
\end{acknowledgments}
\longcomment{
    NOTE:
    Zotero exports English papers as "langauge = {en}", which causes problems for babel.
    As a temporary fix, the affected lines have been manually changed to "language = {english}".
}
\bibliography{references}
\end{document}


\section{Supplementary Material}
Modelling whether or not an applicant scores above a particular P-GRE cutoff score is an example of a binary response problem. This is fortunately an especially common problem in statistical analysis and the techniques employed in this study, i.e., logistic regression and conditional inference forests, are both robust and well-tested. The two techniques differ in their methodology. Whereas logistic regression builds on probabilistic modelling, the forest relaxes interpretation in favor of maximizing predictive accuracy. In the following we provide a short review of the mathematical details of these techniques.
\subsection{Logistic regression modelling}\label{sec:logistic_regression}
Logistic regression is perhaps the most commonly used binary response modelling technique. The main idea is to model the probability of a binary event using linear regression, but since ordinary linear regression fails to adequately restrict the probability range to \([0,1]\), the range is extended to \(\mathbb{R}\) using a log-odds, or ``logit'', transformation. The logit transformation and its inverse, the ``logistic'' transformation, are defined for \(x\in\mathbb{R}\) and \(p\in[0,1]\) by
\begin{subequations}
\begin{align}
    \label{eq:logit_transformation}
    x&=\logit(p)=\ln\big(\odds(p)\big)=\ln\bigg(\frac{p}{1-p}\bigg)\\
    \label{eq:logistic_transformation}
    p&=\logit^{-1}(x)=\frac{1}{1+e^{-x}}
\end{align}
\end{subequations}
While not as restrictive as linear regression, logistic regression does have some limitations \cite{devore_modern_2012}. Most importantly, the input features should not be multicollinear, i.e., no pair of features should be linearly dependent, and each observation must be independent. In addition, logistic regression also requires complete case analysis.

The framework of logistic regression supports both continuous and categorical (discrete) features. However, whereas the continuous features can be used out-of-the-box, the categorical features must be numerically encoded. There is no universal encoding scheme that works well in all cases, but the most common scheme (and the one used in this study) is called \emph{one-hot} encoding. For binary variables, this simply means encoding one outcome as 0 and the other as 1. For variables with \(m>2\) distinct categories, the original variable is replaced by \(m-1\) binary variables. For example, a 3-leveled feature \(x\in\{a,b,c\}\) is replaced by 2 binary features \(x_1\) and \(x_2\) such that
\[x\in\{a,b,c\}\iff(x_1,x_2)\in\big\{(0,0),(1,0),(0,1)\big\}.\]

The setup for logistic regression requires a random Bernoulli-distributed binary outcome \(y\in\{0,1\}\) with a mean of \(p\). The logistic regression model then imposes a linear regression model on \(\logit(p)\) as follows:
\begin{equation}
    \label{eq:logistic_regression}
    \logit p=\beta_0+x_1\beta_1+\cdots x_k\beta_k+\epsilon,
\end{equation}
Here, the \(\beta\)'s are regression coefficients, the \(x\)'s are input features (categorical features one-hot encoded) and \(\epsilon\) is an error term. It is standard practice to collect the \(\beta\)'s and \(x\)'s in vectors \(\boldsymbol{\beta}=[\beta_0,\beta_1,\ldots,\beta_k]^T\) and \(\vb{x}=[1,x_1,x_2,\ldots,x_k]^T\) such that the linear combination can be written as \(\boldsymbol{\beta}^T\vb{x}\). By exponentiating equation \eqref{eq:logistic_regression}, we arrive at the odds equation:
\begin{equation}
    \label{eq:logistic_regression_odds}
    \odds(p)=e^{\boldsymbol{\beta}^T\vb{x}+\epsilon}
\end{equation}
Unlike linear regression, there is no closed form expression for determining logistic regression coefficients. These are determined by an iterative algorithm based on maximum likelihood estimation, which in our study is handled by the \texttt{glm} function in \texttt{R} \cite{friedman_regularization_2010}.
\subsubsection{Interpreting logistic regression coefficients}
A major benefit of logistic regression modelling is the interpretability of its regression coefficients. When \(x_j\) increases by 1 unit, the odds change by a factor of \(\exp(\beta_j)\) called the odds ratio of \(p\) with respect to \(x_j\):
\begin{equation}
    \label{eq:odds_ratio}
    \text{OR}(p;x_j)=\frac{\odds(p;x_j+1)}{\odds(p;x_j)}=e^{\beta_j}.
\end{equation}
The interpretation of the odds ratio depends on whether \(x_j\) is continuous or categorical. For continuous features, the change is associated with a unit increase in \(x_j\). For binary features, the change is associated with a switch in \(x_j\) from category 0 to category 1. The interpretation of \(\exp(\beta_0)\) is slightly more involved; it is the base odds when all continuous features are 0 and all categorical features are in the default category, i.e., when all \(x_j=0\). Odds ratios below 1 are inverted so that \(1/\text{OR}(x_j)\) is the odds ratio associated with a unit decrease in \(x_j\) or a switch in \(x_j\) from category 1 to category 0.

Because categorical features with more than two levels are divided into binaries (one-hot encoding), the odds ratio for each binary is associated with a switch from the reference category to the category corresponding with the binary. For the \(x\in\{a,b,c\}\) example, the odds ratio of \(x_1\) would be associated with switching \(x\) from \(a\) to \(b\). The reader should note how the odds ratios of categorical features are always in relation to the reference level. Hence, selecting another reference level may highlight (or hide) different aspects of the relationship between categories. While there is no standard practice for identifying suitable reference levels, typically the ``lowest'' category is chosen for ordinal categories and the ``ordinary'' category is chosen when there is a natural ``normal state''. See Sperandei \cite{sperandei_understanding_2014} (2014) for additional strategies.

A problem that is especially important for models with a mixture of continuous and categorical features is that of uneven scales of the continuous variables. A feature of order \(10^5\) will have an incredibly small odds ratio if all other features are of order 1, regardless of the feature's actual impact in the model. To counteract these problems, it is common to \emph{standardize} the continuous features, that is, center the mean about 0 and normalize the variance to 1. For standardized features, the odds ratio is associated with an increase in the original feature by one standard deviation.
\subsubsection{Statistical significance}
Alongside the regression coefficients, the \texttt{glm} function provides the corresponding \(z\)-score statistics from which one can determine the confidence interval and \(p\)-value of the coefficients (via a Wald test). Typically, a \(p\)-value is considered significant if it is below a significance level of \(\alpha=0.05\). However, in order to avoid multiple comparisons problems, the \(p\)-value is adjusted according to the Bonferroni correction. For a logistic regression model with \(N\) features, i.e., with \(N\) \(\beta_j\)'s in \eqref{eq:logistic_regression}, the Bonferroni-adjusted \(p\)-value, \(\tilde{p}=pN\), is compared to \(\alpha\).

Using \eqref{eq:odds_ratio} the odds ratios and their corresponding confidence intervals can be computed. In terms of the odds ratios, a feature is statistically significant under the Bonferroni correction if a ratio of 1 lies outside the \((1-\alpha)\%\) confidence interval of \(\tilde{p}\).
\subsubsection{Caveats with logistic regression}
Due to its assumptions and requirements, logistic regression suffers from a number of drawbacks. For example, requiring complete case analysis forces the investigator to intervene with any missing data by means of imputation or simply ignoring it. Although either choice will effect the analysis, Nissen et al. argues in favor of imputation for PER \cite{nissen_missing_2019}. Admittedly not unique to logistic regression, the framework is also very sensitive to imbalanced categorical features. For categorical features with low-frequency categories, the maximum likelihood methods may not converge or may yield absurd results such as infinitely large confidence intervals. Indeed, many of the features used in this study had to be preprocessed in order to counteract the sensitivity of logistic regression.
\subsection{Supervised classification analysis}
Given a data set with input-output pairs \((x,y)\), statistical analyses typically revolve around a map \(f\) from \(x\) to \(y\): \(y=f(x)\). In logistic regression, the map is given by the odds equation \eqref{eq:logistic_regression_odds}, whose parametric details, i.e., the values of \(\beta_j\), are inferred using the data set. Much of the problems faced by logistic regression stems from its probabilistic interpretation. At the cost of loosing this interpretation, supervised machine learning (SML) techniques ignore cause and effect, instead focusing on statistical relationships for the benefit of flexibility and accurate predictions. That is, instead of fixing the shape of \(f\) and inferring \(\beta_j\), SML attempts to construct a map \(f\) based entirely on the data set. Hence, SML is not concerned with whether there is a physical mechanism that connects \(x\) and \(y\), only the search for \(f\). Depending on whether \(y\) is continuous or discrete, SML is categorized as ``regression'' or ``classification'' respectively. A classification map \(f\) is commonly referred to as a \emph{classifier} in the sense that it classifies inputs into discrete output classes. There are many algorithms for producing a map \(f\); this study employs the \emph{conditional inference forest} algorithm, which is a variant of the ordinary \emph{random forest} algorithm \cite{breiman_random_2001, hothorn_unbiased_2006}. A benefit of these forest algorithms comes from their construction, which allows for a wide variety of data structures without requiring data preprocessing.

That supervised classification is focused on discrete outputs is somewhat of a red herring. A number of classification algorithms, including the forest algorithms, provide a way to predict the probability \(p\) of the output being in a particular class. The class assigned to the inputs is then selected based on the most probable class. However, while the number provided by the algorithm behaves like a probability (\(0\leq p\leq1\)), it is not required to be. Hence, the best prediction is not necessarily the class for which the ``probability number'' is the greatest. For our case, this means that the P-GRE score of an applicant is predicted as \emph{above} the cutoff if the probability number is greater than a certain \emph{probability threshold}, which is chosen to maximize the performance of the classifier. Depending on the data set and algorithm, adjusting the probability threshold can significantly change the performance of a classifier. Indeed, adjusting the threshold was recently used in a study of unbalanced physics course outcomes to drastically improve the performance of a random forest classifier \cite{devore_extending_2020}.
\subsubsection{Estimating the performance of a classifier}
The main way of assessing the performance of a classifier is to test ``how well'' it is able to predict the output of a known data set using so-called \emph{performance metrics}. In order to avoid overestimating performance, the standard practice involves splitting the data set at random into so-called \emph{training} and \emph{testing} sets. After learning the training set, i.e., constructing \(f\), the classifier is tested by comparing the true and predicted outputs of the testing set. Some performance metrics are presented below.

Dividing the data is a sensitive operation as it may lead to unreasonably high or low scores by chance. To avoid incorrectly assessing a classifier, the ``true'' performance is estimated by \emph{cross-validation}. The principle behind the most common algorithm, the \(K\)-fold, is sketched in figure \ref{fig:Kfold_cross_validation}. \(K\)-fold divides the (complete) data set into \(K\) disjoint folds (samples) of roughly the same size. For each fold \(i=1,2,\ldots,K\), the classifier is first trained on the other \(K-1\) folds, then tested on fold \(i\). The performance metrics are then averaged over all \(K\) iterations. The corresponding standard errors are used to assess the quality of the cross-validation estimates.
\begin{figure}
    \centering
    \begin{tikzpicture}
    \foreach [evaluate] \x in {1,2,3,4,5} {
        \foreach[evaluate={\color = \x==\y ? "cyan" : "white"}] \y in {1,2,3,4,5} {
            \node at ({-0.6+1.2*\x}, {-0.1-0.5*\y}) [rectangle, draw, fill=\color] {\x};
        }
    }
    \draw[->,>=stealth] (0,0) -- ({+5*1.2+0.1},0) node [pos=0.5, above] {folds};
    \draw[->,>=stealth] (0,0) -- (0,{-6*0.5-0.1}) node [pos=0.5, above, rotate=90] {iterations};
    \end{tikzpicture}
    \caption{A sketch of the \(K\)-fold cross-validation algorithm using \(K=5\). The original data set is divided into \(K\) disjoint subsets. For each iteration, the performance metrics are trained on the empty folds and tested on the colored fold.}
    \label{fig:Kfold_cross_validation}
\end{figure}
\subsubsection{Performance metrics}\label{sec:performance_metrics}
For binary outcomes, it is customary to label the output classes \emph{Positive} and \emph{Negative}, then label the predictions \emph{True} or \emph{False} depending on whether the prediction was correct. Using this terminology, a correct prediction of \emph{Positive} is referred to as a \emph{True Positive} (TP), while an incorrect prediction of \emph{Positive} is referred to as a \emph{False Positive} (FP). Similar terminology is attributed to \emph{True} and \emph{False Negative} (TN and FN).

A common tool for categorizing the predictions of a classifier, and one that gives rise to several performance metrics, is the confusion matrix. It is a particular type of two-way contingency table where the columns describe the actual outcomes and the rows describe the predicted outcomes. For a binary classification problem, the confusion matrix is a \(2\times2\) matrix that arranges the number of TP, FP, TN and FN in a testing set. Table \ref{tab:confusion_matrix} shows the confusion matrix used in the present study. The numbers on the diagonal (TP and TN) represent correctly classified labels, while the numbers on the off-diagonal (FP and FN) represent incorrectly classified labels. FP and FN are also referred to as Type I (false alarm) and Type II (miss) errors in the literature. Note that since the confusion matrix is dependent on the predictions of the classifier, it is also dependent on the chosen probability threshold of the classification, and thus all performance metrics derived from the confusion matrix will also be dependent on the threshold.
\begin{table}[h]
    \centering
    \caption{The confusion matrix for the binary \emph{above}/\emph{below} classification problem focused on in this study. The actual outcomes are the binary outputs in the testing set, while the predicted outcomes are the predictions of the classifier when given the corresponding inputs of the testing set.}
    \label{tab:confusion_matrix}
    {\renewcommand{\arraystretch}{1.9}
    \settowidth\rotheadsize{\theadfont Outcomes}
    \begin{tabular}{cc|cc}
        & \multicolumn{1}{c}{} & \multicolumn{2}{c}{Actual Outcomes}                      \\[-2mm]
        & \multicolumn{1}{c}{} & \emph{above}               & \emph{below}                \\\cline{3-4}
    \multirow{3}{*}[2mm]{\rothead{Predicted\\Outcomes}}
        & \emph{above}         & \makecell{True\\Positives} & \makecell{False\\Positives} \\
        & \emph{below}         & \makecell{False\\Negatives} & \makecell{True\\Negatives}
    \end{tabular}}
\end{table}

The most common and straightforward performance metric is the accuracy score, which is defined as the portion of correctly classified labels. In terms of the confusion matrix, accuracy is given by
\begin{equation}
\accuracy=\frac{\TP+\TN}{\TP+\FN+\FP+\FN}
\end{equation}
The accuracy score ranges between 0 and 1, where a score of 1 signifies a perfect classifier. Accuracy is usually denoted by a percentage, i.e., 80\% instead of 0.8. While easy to interpret, the accuracy score is very sensitive to unbalanced classes because it does not distinguish between correctly classifying a \emph{Positive} and a \emph{Negative} label. For instance, if 80\% of outputs are \emph{Positive}, a naive classifier predicting \emph{Positive} regardless of the inputs would then have an accuracy score of 80\%. For this reason, accuracy should be considered relative to the binary output's class imbalance.

Another commonly used metric is the AUC score, or the area under the receiver operating characteristic (ROC) curve. A sketch of typical ROC curves is shown in figure \ref{fig:ROC_sketch}. The ROC curve graphs the true positive rate (TPR) against the false positive rate (FPR), also known as \emph{sensitivity} and \emph{fall-out} respectively. Both TPR and FPR range from 0 to 1 like the accuracy score. Sensitivity is a measure of a classifier's ability to correctly identify \emph{Positive} labels, while fall-out is the rate of Type II errors. In terms of the confusion matrix, TPR and FPR are given by
\begin{equation}\label{eq:TPR_and_FPR}
\TPR=\frac{\TP}{\TP+\FN}\qand \FPR=\frac{\FP}{\FP+\TN}.
\end{equation}
As the probability threshold is increased from 0 to 1, the ROC-coordinates (FPR,TPR) trace out a curve within the region \([0,1]\times[0,1]\). Since the AUC score is the area under the ROC curve, the dependence on the probability threshold is lost. There is also a probabilistic interpretation of the AUC score; it is the probability that a classifier ranks a random \emph{Positive} label higher than a random \emph{Negative} label \cite{fawcett_introduction_2006}. That is, AUC measures how well a classifier is able to distinguish between \emph{Positive} and \emph{Negative} outcomes. Like accuracy, AUC is also bound by 0 and 1, where 1 is a perfect score.
\begin{figure}
    \centering
    \begin{tikzpicture}
        \draw[->, >=stealth] (-0.2,+0.0) -- (+5.0,+0.0) node [pos=0.5, below] {False positive rate};
        \draw[->, >=stealth] (+0.0,-0.2) -- (+0.0,+5.0) node [pos=0.5, above, rotate=90] {True positive rate};
        \draw[dashed] (0,0) -- (4.5,4.5);
        \draw (0,4.5) -- (4.5,4.5);
        \draw[thick] (-0.1,4.5) -- (+0.1,4.5);
        \draw[thick] (4.5,-0.1) -- (4.5,+0.1);
        \draw node at (-0.3,-0.3) {0};
        \draw node at (-0.3,+4.5) {1};
        \draw node at (+4.5,-0.3) {1};
        \draw[blue] (0,0) .. controls (1.3,3.0) and (3.4,4.0) .. (4.5,4.5);
        \draw[red]  (0,0) .. controls (0.0,4.5) and (2.0,4.4) .. (4.5,4.5);
        \node[draw=black, rounded corners=2pt, below=1mm] at (current bounding box.east) {
            \begin{tabular}{@{}r@{ }l@{}}
                \multicolumn{2}{c}{\scriptsize{Classifier:}}\\[-1mm]
                \raisebox{2pt}{\tikz{\draw[black]        (0,0) -- (3mm,0);}} & \scriptsize{perfect}        \\[-1.2mm]
                \raisebox{2pt}{\tikz{\draw[red]          (0,0) -- (3mm,0);}} & \scriptsize{excellent}      \\[-1.2mm]
                \raisebox{2pt}{\tikz{\draw[blue]         (0,0) -- (3mm,0);}} & \scriptsize{poor}           \\[-1.2mm]
                \raisebox{2pt}{\tikz{\draw[black,dashed] (0,0) -- (3mm,0);}} & \scriptsize{random chance}
            \end{tabular}
        };
    \end{tikzpicture}
    \caption{Sketch of typical ROC curves (axes defined in \eqref{eq:TPR_and_FPR}). Each point on the ROC curve corresponds to a particular classification probability threshold \(p\in[0,1]\). The closer an ROC curve resembles random chance (dashed line), the worse the corresponding classifier is.}
    \label{fig:ROC_sketch}
\end{figure}

In general, there is no agreement on what constitutes ``good'' performance. For accuracy, we consider a score equivalent to the class imbalance to be naive in the sense that the classifier performs as well as a majority vote. A classifier is considered not naive if the class imbalance is outside the \(K\)-fold standard error bounds of the estimated accuracy score. For AUC, a classifier that is unable to distinguish between the output classes would score \(\approx0.5\) (equivalent to random guessing). Because no universal standard exists, AUC is most commonly used for comparing classifiers, and is generally preferred over accuracy for single-number comparisons \cite{bradley_use_1997}. Nevertheless, Hosmer et al. suggests this general rule: \(0.7\leq\AUC<0.8\) is acceptable, \(0.8\leq\AUC<0.9\) is excellent, and \(0.9\leq\AUC\) is outstanding \cite{hosmer_applied_2000}.
\subsubsection{Conditional inference trees}\label{sec:conditional_inference_forest}
Before continuing to the forest algorithms, its instructive to first consider the trees from which they are built. Like the name implies, decision trees are an algorithmic approach to decision-making, which in the context of classification analysis is related to predicting output classes. Decision trees, or simply trees, are named after their graphical representation of connected nodes resembling an upside-down family tree diagram as depicted in figure \ref{fig:decision_tree_sketch}. Each node represents a test regarding a particular input feature and each tree branch represents a decision. Depending on the result of a test, a specific connection (branch) is followed to the next node. This process is repeated until the terminal nodes, or tree leaves, are reached and a decision is made. Importantly, several leaves can result in the same decision. In general, a feature is more important the closer it is to the top (root) node. Here, the definition of ``\emph{important}'' depends on the algorithm constructing the decision tree.
\begin{figure}[h]
    \centering
    \begin{tikzpicture}[fill=cyan]
        \draw node at (+0.0,+0-0) [circle,draw]      (a) {\scriptsize{\(x\geq0\)}};
        \draw node at (+0.8,-1.5) [circle,draw,fill] (b) {\scriptsize{\(C\)}};
        \draw node at (-0.8,-1.5) [circle,draw]      (c) {\scriptsize{\(y\geq0\)}};
        \draw node at (-1.6,-3.0) [circle,draw,fill] (d) {\scriptsize{\(A\)}};
        \draw node at (+0.0,-3.0) [circle,draw,fill] (e) {\scriptsize{\(B\)}};
        \draw (c) -- (a) -- (b);
        \draw (d) -- (c) -- (e);
    \end{tikzpicture}
    \caption{A sketch of a decision tree with three leaves that is dependent on two continuous variables \(x\) and \(y\). Depending on the value of \(x\), the decision is either \(C\) or it continues to the left node. If it continues, the final decision is \(A\) or \(B\) depending on \(y\).}
    \label{fig:decision_tree_sketch}
\end{figure}

The decision tree algorithm employed in this study is the conditional inference tree algorithm, which was introduced by Hothorn et al. in 2006 \cite{hothorn_unbiased_2006}. This particular variant solves issues relating to overfitting, the problem of SML algorithms learning the random errors of a training set, and selection bias towards features with many possible splits (node tests) \cite{boulesteix_random_2012,strobl_bias_2007}. While constructing a tree node, the algorithm select the most important feature using a procedure known as \emph{permutation testing}, which essentially finds the feature that is most associated with the response. Starting from the root, the algorithm constructs the nodes one-by-one until no features are left or a \emph{stopping criterion} is reached. In this study, the stopping criterion is based on the significance level \(\alpha=0.05\) at which the hypothesis ``\emph{the distribution of the response is independent of the remaining features}'' is rejected. See ``\emph{Variable selection and stopping criteria}'' in section 3 of Hothorn et al. \cite{hothorn_unbiased_2006} for more details.
\subsubsection{Conditional inference forests}
The conditional inference forest (CIF) algorithm is modern variant of the popular random forest algorithm developed by Breiman \cite{breiman_random_2001}. A random forest consists of an ensemble of decision trees, each of which is an independent classifier. The algorithm (Breiman variant) applies a combination of ``bagging'' (bootstrap aggregation) and random subspace methods: each tree is trained on a random selection of the features with a bootstrap sample of the training set. Predictions of the forest are then based on the aggregate results of the ensemble: the probability of an output class is evaluated by a majority vote among the trees. Despite its popularity, the random forest algorithm has several drawbacks \cite{strobl_bias_2007}, most notably its biased variable selection. For a discussion on the ways in which the biased variable selection affects data similar to data used in the present study, see the supplementary material of Young et al. \cite{young_identifying_2019} (2019). These issues are resolved by the CIF however, which is similar to the random forest, but built with conditional inference trees and constructed by subsampling instead of bagging. In the present study, the CIF algorithm is handled by the \texttt{party} package in \texttt{R}, written by Hothorn et al. \cite{hothorn_survival_2006,strobl_bias_2007,strobl_conditional_2008}.

Many SML algorithms depend on so-called ``\emph{hyperparameters}'', which are adjusted to optimize performance. In contrast to ``ordinary parameters'' that are determined during the construction of \(f\) (e.g., \(\beta_j\) in logistic regression), hyperparameters specify details regarding the construction algorithm itself. There are primarily two hyperparameters to consider for the CIF algorithm: the number of trees per forest, and the number of features used to construct a tree. We follow the standard suggestion for classification problems of growing a tree based on \(\approx\sqrt{k}\) out of \(k\) features (see e.g., Svetnik et al. \cite{svetnik_random_2003}). As for the forest, increasing the number of trees is an example of diminishing returns as the performance stabilizes after a certain number of trees has been added. In our case, minimal random errors are observed with 200 trees (Svetnik et al. also found that about 100 trees, depending on the size of the data, are sufficient for stable results). Other CIF hyperparameters have been left to the defaults of \texttt{party}.

Due to their construction, CIFs are incredibly versatile: data sets can have both continuous and categorical data; categorical features do not need to be numerically encoded, and continuous features do not need to be standardized. The CIF framework was also extended by Hapfelmeier et al. to handle missing values \cite{hapfelmeier_new_2014}. In addition to its flexibility, CIFs have built-in techniques for estimating the relative importances of input features by averaging the impact (across all trees) of a removing each feature (separately) according to an importance measure. Prediction accuracy have been used extensively for this purpose, but Janitza et al. argues in favor of the AUC score as it is more robust towards imbalanced data sets \cite{janitza_auc-based_2013}.
\bibliographystyle{apsper}
\bibliography{references}